\documentclass[11pt]{article}
\usepackage{geometry} 
\usepackage{graphicx}
\usepackage{amssymb}
\usepackage{epstopdf}
\usepackage{lmodern}
\usepackage{palatino}
\usepackage{amsmath}
\usepackage{mciteplus}
\usepackage{subfigure}
\usepackage{float}
\usepackage [english]{babel}

\usepackage{xcolor}
\definecolor{lcolor}{rgb}{0.,0.0,0.}
\definecolor{citcolor}{rgb}{0,0.,0.5}
\usepackage[breaklinks,colorlinks,urlcolor=blue,citecolor=blue,linkcolor=blue]{hyperref}

\newcommand{\beq}{\begin{equation}}
\newcommand{\eeq}{\end{equation}}
\newcommand{\bea}{\begin{eqnarray}}
\newcommand{\eea}{\end{eqnarray}}

\newcommand{\bem}{\begin{multline}}
\newcommand{\eem}{\end{multline}}
\newcommand{\beg}{\begin{gather}}
\newcommand{\eeg}{\end{gather}}

\def\eq#1{{Eq.~(\ref{#1})}}

\newcommand{\ben}{\begin{eqnarray*}}
\newcommand{\een}{\end{eqnarray*}}

\title{Ultra-forward particle production from CGC+Lund fragmentation}
\author{
{Javier L. Albacete$^a$, Pablo Guerrero Rodr\'{\i}guez$^a$, Yasushi Nara$^b$}\\[0.2cm]  {\it \small $^a$CAFPE and Departamento de F\'isica Te\'orica y del Cosmos,  Universidad de Granada}\\ {\it \small E-18071 Campus de Fuentenueva, Granada, Spain.} \\{\it \small $^b$Akita International University, Yuwa, Akita-city 010-1292, Japan.} \\[0.1cm] {\texttt{ \small albacete@ugr.es, pgr@ugr.es, nara@aiu.ac.jp}}
}
\date{}
\begin{document}
\maketitle

\begin{abstract}
We present an analysis of data on single inclusive pion production measured by the LHCf collaboration in high-energy proton-proton and proton-nucleus at ultra-forward rapidities, $8.8 \leq\! y\leq \!10.8$. We also analyse forward RHIC data for calibration purposes. Our analysis relies on the use of a Monte Carlo event generator that combines a perturbative description of the elementary scattering process at partonic level based on the hybrid formalism of the Color Glass Condensate with an implementation of hadronization in the framework of the Lund string fragmentation model. This procedure allows us to reach values of the momenta of the produced particles as low as detected experimentally $p_t\sim0.1$ GeV.
We achieve a good description of single inclusive spectra of charged particles and neutral pions at RHIC and the LHC respectively, and nuclear modification factors for proton-lead collisions at the LHC. Our results add evidence to the idea that particle production in the domain of very small Bjorken-$x$ is dominated by the saturation effects encoded in the unintegrated gluon distribution of the target. Being forward particle production of key importance in the development of air showers, we stress that this approach allows for a theoretically controlled extrapolation of our results to the scale of ultra-high energy cosmic rays, thus serving as starting point for future works on this topic.
\end{abstract}

\section{Introduction}\label{introduction}

The detection and analysis of particle production in collision processes in the very forward region allows to study wave functions of the colliding objects in extreme limits of the phase space. At partonic level these collisions can be interpreted as mediated by a highly energetic valence quark from the projectile scattering off a soft or wee parton (typically a gluon) of the target hadron. After hadronization, the produced particles fly very close to the beam pipe until reaching the forward detectors. The Bjorken-$x$ values proben in the scattering process can be estimated to be $x_{p,t}\approx (p_t/\sqrt{s})e^{\pm y}$ (more details below), where $\sqrt{s}$, $p_t$ and $y$ refer to the collision energy, transverse momentum and rapidity of the produced particle. In the kinematic range covered by LHCf one has $\sqrt{s}\!=\!7$ TeV, $p_t\lesssim\! 1$ GeV and $8.8 \leq\! y\leq \!10.8$, yielding the following Bjorken-$x$ values for projectile and target: $x_p\sim 10^{-1}\sim1$ and $x_t\sim 10^{-8}\sim 10^{-9}$. The latter values of Bjorken-$x$ are the smallest ever accessed in collider experiments. In this work we analyse forward production data in proton-proton and proton-nucleus collisions at RHIC and the LHCf in terms of saturation physics and non-linear small-$x$ QCD evolution equations. In terms of evolution rapidity $Y=\ln(x_0/x)$, the simultaneous description of RHIC (where the typical Bjorken-$x$ values of the target are $\approx 10^{-3}$) and LHCf data implies testing such equations over a rapidity interval $\Delta Y \sim 14$.
%In the case of forward neutral pion production in proton-proton collisions at LHC, this analysis was already performed by Yasushi Nara et al. using the same approach featured in this paper. Their results were presented in \cite{Deng:2014vda}.

From the theoretical point of view it is well stablished that at small values of Bjorken-$x$ QCD enters a new regime governed by high gluon densities and the coherent dynamics of the soft color fields, consistently accounted for by the non-linear renormalization group equations of QCD, the BK and B-JIMWLK equations~\cite{Jalilian-Marian:1997gr,Jalilian-Marian:1997dw,Kovner:2000pt,Weigert:2000gi,Iancu:2000hn,Ferreiro:2001qy,Balitsky:1996ub,Kovchegov:1999yj}. The presence of non-linear terms in the small-$x$ evolution equations limits the growth rate of gluon number densities for modes of transverse momentum smaller than the saturation scale $Q_s(x)$. This novel, semi-hard dynamical scale marks the onset of non-linear corrections in QCD evolution and leads to distinctive dynamical effects that, as we argue below, reflect in the ultra-forward particle spectra measured by the LHCf collaboration. In addition to theoretical arguments, there is by now an abundant corpus of phenomenological works that provides compelling evidence for the presence of saturation effects in currently available experimental data in a variety of collision systems: electron-proton, proton-proton and heavy ion collisions. For a review, see e.g. \cite{Albacete:2014fwa}. 

LHC and RHIC forward particle production data have been analysed in a series of previous works~\cite{Albacete:2012xq, Albacete:2010bs,Stasto:2013cha} in the framework of the Color Glass Condensate (CGC) effective theory for high energy QCD scattering (see e.g~\cite{Weigert:2005us,Gelis:2010nm} for a review). Same as in this work, the above mentioned works relied on the use of the hybrid factorization, well suited for the description of forward production processes or equivalently  {\it dilute-dense} scattering, thus referred due to the strong asymmetry in the Bjorken-$x$ values probed in the projectile and target. 
The main novelty of this work with respect to previous ones is the treatment of the hadronization process, which we describe in the framework of Lund string fragmentation.  This procedure allows us to reach values of the momenta of the produced particles as low as detected experimentally $p_t\sim0.1$ GeV and, therefore, opens the possibility of describing particle multiplicities. In turn, the description of the hadronization process in terms of fragmentation functions used in previous works limits their applicability, by construction, to perturbatively large values of transverse momentum of the produced particle, $p_t\gtrsim 1$ GeV. 
Further, we also take into account the possibility of multiple, simultaneous scatterings of different valence quarks with the dense glue of the target. The three main ingredients of our set up, namely perturbative partonic production as given by the hybrid formalism, Lund fragmentation and multiple scatterings are embedded in a Monte Carlo event generator. The Monte Carlo Code used here was first developed by one of us for the description of ultra-forward pion production in proton-proton collisions at the LHC\cite{Deng:2014vda}. We shall extend it here to the case of proton-nucleus collisions and the study of the measured nuclear modification factors.

\section{Set Up}\label{setup}
A highly asymmetric collision in terms of the proben $x$-values is also strongly asymmetric in the density of the colliding objects. In our case of study, the dilute system is given by the ensemble of valence partons in the proton projectile, which scatters off the highly dense, coherent glue in the target, be it a proton or nucleus. In the hybrid formalism the large-$x$ degrees of freedom of the proton are described in terms of usual parton distribution functions (PDFs) of collinear factorization which satisfy the momentum sum rule exactly and which exhibit a scale dependence given by the DGLAP evolution equations. On the other hand, the small-$x$ glue of the nucleus is described in terms of its unintegrated gluon distribution (uGD). The hybrid formalism was first proposed in \cite{Dumitru:2005gt} and, at partonic level, the cross section for quark or gluon production in the scattering off a dense gluonic target reads:

	\begin{equation}\label{DHJ}
\frac{d\sigma^{h_1h_2\to (q/g)X}}{dy\,d^2k}=\frac{K}{(2\pi)^2}\frac{\sigma_0}{2}\,x_1f_{(q/g)/h_1},
(x_1,\mu^{2})N_{(F/A),h_2}\left(x_2,k_t\right). 
\end{equation}
where $f_{(q/g),h_1}(x_1,\mu^2)$ is the PDF of quarks or gluons in the projectile $h_1$ evaluated at the scale $\mu$, while $N_{(F/A),h_2}\left(x_2,k_t\right)$ refers to the uGD of the target  in either the fundamental (F, for quarks) or adjoint (A, for gluons) representation. The uGDs are related to the dipole scattering amplitude in coordinate space via a Fourier transform:  

\begin{equation}
N_{(F/A)}(x,k_t)=\int d^2{\bf r}\;
e^{-i{\bf k_t}\cdot{\bf r}}\left[1-\mathcal{N}_{(F/A)}(x,r)\right].
\label{phihyb}
\end{equation}
We shall take the parametrization of the dipole scattering amplitude $\mathcal{N}(x,r)$ from the AAMQS fits to data on the structure functions measured in e+p scattering at HERA\cite{Albacete:2009fh,Albacete:2010sy}. The main dynamical input in those fits is the running coupling BK equation for the description of the $x$-dependence of the dipole amplitudes\cite{Kovchegov:2006vj,Balitsky:2006wa,Albacete:2007yr}. The fit parameters are mostly related to the initial conditions for the evolution, set at the initial Bjorken-$x$ value $x_0\!=\!10^{-2}$. In the AAMQS fits they were chosen in the following form:
	\begin{equation}\label{Init}
		\mathcal{N}_{F}(x_{0},r)=1-\exp{\left[-\frac{(r^{2}Q_{s0}^{2})^{\gamma}}{4}\log{\left(\frac{1}{\Lambda r}+e\right)}\right]}.
	\end{equation}
The AAMQS fits provide a well constrained parametrization of the proton uGD. Similar to what has been done in previous works\cite{Albacete:2010ad}, we build the uGD of a nuclear target (lead or gold in our case), by simply rescaling the value of the initial saturation scale in \eq{Init} by the \textit{oomph} factor: $Q^2_{s0,nucleus}=A^{1/3}Q_{s0,proton}^2$, where $A$ is the mass number of the target nucleus. The corresponding dipole scattering amplitudes in the adjoint representation are given, in the large-$N_c$ limit, by 	$\mathcal{N}_A\!=\! 2\mathcal{N}_F-\mathcal{N}^2_F$. In this work we shall use the AAMQS sets corresponding to $\gamma\!=\!1.101$, $Q^2_{s0}\!=\!0.157$ GeV$^2$ and $\gamma\!=\!1.119$, $Q^2_{s0}\!=\!0.168$ GeV$^2$. It turns out that the results for LHCf kinematics are very little sensitive to this particular choice and other AAMQS sets yield a very similar description of the data. 

As for the proton PDFs, we shall take the CTEQ6 leading order set \cite{Pumplin:2002vw} with a default factorization scale $\mu=\text{max}\{ k_{t}, Q_{s} \}$, where $k_{t}$ is the transverse momentum acquired by the incoming parton as it multiply scatters the soft glue of the target. This choice ensures that primary partonic production can be described by means of perturbative tools. We cannot exclude that part of primary particle production could be of genuinely non-perturbative origin --specially for very small transverse momenta of the produced pions-- and hence, not amenable to description in terms of \eq{DHJ}. However, the good description of the data reported below makes us confident that the main dynamical features of the process studied here are well accounted for by our approach.
When applied to LHCf kinematics, our ansatz for the factorization scale ensures that it always falls into the perturbative domain $\mu\gtrsim 1$ GeV, since the saturation scale at the LHC ultra-forward region is perturbatively large: $Q_s(x\sim10^{-8})\gtrsim 1$ GeV, both for proton and lead targets. Such is not the case in RHIC kinematics, where the saturation scale is considerably smaller and closer to its initial values $Q_{s0}^2\sim 0.2$ GeV. In the latter case, we impose a momentum cutoff on the exchanged transverse momentum $k_{t,\text{min}}=1\,\text{GeV}$. However, this cutoff is not necessary at the LHCf or, in other words, our results are insensitive to its precise value, as the scattering is dominated by higher transverse momenta, of the order of the saturation scale of the target $k_t\sim Q_s(x)\gtrsim 1$ GeV.

%Finally, the factor $\sigma_0$ in \eq{DHJ} results from the integration over impact parameter implicit in \eq{phihyb}. In the mean field approach treatment of the target geometry --proton or nucleus-- that we shall adopt here, it carries the meaning of its average transverse size.
%For proton, its value can be taken from the AAMQS fits where it was one of the free fit parameters ($\sigma_{0}/2=16.5\,\text{mb}$), whereas for nuclei we simply scale the corresponding value for proton by the geometric factor scale $A^{2/3}$.%
Finally, the factor $\sigma_0$ in \eq{DHJ} results from the integration over impact parameter implicit in \eq{phihyb}. In the mean field approach treatment of the target geometry --proton or nucleus-- that we shall adopt here, it carries the meaning of the average transverse size of the proton. Its value can be taken from the AAMQS fits where it was one of the free fit parameters ($\sigma_{0}/2=16.5\,\text{mb}$). The $K$-factor in \eq{DHJ} is not the result of any calculation. It has been added by hand to account for higher order corrections, possible non-perturbative effects, etc. In practice, we use it to adjust the normalization of theoretical curves to experimental data in phenomenological works. In an ideal situation it should be equal to unity.

%Recopilar y citar algunos achievements de la DHJ	
%The Dumitru-Hayashigaki-Jalilian-Marian (DHJ from now on) formula (\ref{DHJ}) has been used to reproduce the forward charged particle spectra in p+p collisions 
%(...), d+Au collisions (...,\cite{Albacete:2010bs},\cite{Hayashigaki:2006ek}) and (...).

%%%%Javi
Before dwelling into more details on our implementation, a comment is in order: the degree of accuracy of the hybrid factorization formula as well as that of the non-linear evolution equations describing the Bjorken-$x$ dependence of the uGD of the target --running coupling BK in our case-- have been considerably improved in the recent past. In particular, next-to-leading order (NLO) corrections to the cross-section \eq{DHJ} have been calculated in \cite{Altinoluk:2014eka, Chirilli:2012jd}. Also, both the BK and B-JIMWLK evolution equations are now known at full NLO accuracy\cite{Balitsky:2008zza,Balitsky:2013fea}. However, it was quickly noticed that the perturbative expansion of hadronic observables and evolution equations at NLO become unstable in certain regions of phase space\cite{Albacete:2012xq,Stasto:2013cha,Lappi:2015fma}, even leading to negative cross sections. Such unphysical behaviour has been identified as due to the increasing importance of double transverse momentum logarithms. Later works showed that the resummation of those collinear logs stabilises the behaviour of the perturbative series\cite{Iancu:2015vea,Lappi:2016fmu}, even allowing a good phenomenological description of electron-proton cross sections measured in HERA\cite{Albacete:2015xza,Iancu:2015joa}. Recently, it has been suggested that the kinematic corrections embodied in the resummation of large collinear logs can be accounted for through an appropriate subtraction of the rapidity divergence in the BK evolution for the target \cite{Ducloue:2016shw}.   

However the notable progress briefly reported above, we shall consider the hybrid formalism \textit{only} at leading logarithmic accuracy (LL) together with LO DGLAP evolution and running coupling BK evolution to describe the scale dependence of the projectile PDF and target uGD respectively. 
Although a full NLO analysis of forward production data would be desirable --as all theoretical tools are now available-- its phenomenological implementation should start by performing a global fit to e+p data at full NLO accuracy in order to obtain the uGD of a proton, which has not been carried out to date. Also, as shown in \cite{Albacete:2012xq,Stasto:2013cha}, NLO effects become increasingly important in the region of high transverse momentum and small to moderate evolution rapidities $Y=\ln(x_0/x)$. In this work we are interested in the opposite kinematic regime of very high evolution rapidities $Y\sim 15$ and small transverse momentum scales, $k_t\lesssim Q_s(x)$. We expect then that the LO implementation of the hybrid factorization captures the main dynamical features of the collision process in the LHCf kinematic regime. This set up could be systematically improved using available theoretical progress, but we leave such task for future works.  

The three main ingredients of our set up, namely primary, perturbative partonic production as given by \eq{DHJ}, Lund fragmentation and multiple scatterings are embedded in a Monte Carlo event generator. 
At partonic level, our Monte Carlo code generates quarks and gluons from $qg\rightarrow q$ and $gg\rightarrow g$ hard scatterings according to \eq{DHJ} along with initial and final state radiation based on DGLAP evolution. Multiple parton scattering is implemented in the eikonal model formalism \cite{Butterworth:1996zw,PhysRevLett.58.303,PhysRevD.40.1436}, where we assume the probability distribution governing the number of independent hard scatterings to be a Poisson of mean $n$, with:
	\begin{equation}\label{MPI}
		n(b,s)=T_{\text{pp}}(b)\sigma_{DHJ}(s).
	\end{equation}
$n$ is the average number of partonic collisions per event. It depends on the invariant mass of the collision $s$ through the integrated cross section $\sigma_{DHJ}$, and on the impact parameter of the collision $b$ through $T_{\text{pp}}$, which is the spatial overlap of the colliding protons obtained as the convolution of two Gaussian functions:
	\begin{equation}\label{Gaus}
		T_{\text{pp}}(b)=\frac{1}{4\pi B}\exp{\left( -\frac{b^{2}}{4B} \right)}.
	\end{equation}
For every event, the impact parameter $b$ is randomly generated in a range between $0\,\text{fm}$ and:
	\begin{equation}\label{bmax}
		b_{max}=\sqrt{\frac{\sigma_{nd}}{\pi}}.
	\end{equation}
which is the radius of a circle of area defined by the cross section of non-diffractive events, $\sigma_{nd}$.
For collisions on nuclear targets, we substitute the target profile by a Gaussian with radius $R^{2}_{\text{A}}=R^{2}_{\text{p}}\,A^{2/3}$. Its convolution with the Gaussian profile of a proton yields:
	\begin{equation}\label{pAgauss}
		T_{\text{pA}}(b)=\frac{A^{2/3}}{\pi R^{2}_{\text{p}}(A^{2/3}+1)}\exp{\left( \frac{-b^{2}}{R^{2}_{\text{p}}(A^{2/3}+1)} \right)}.
	\end{equation}
Which is normalized to $A^{2/3}$.
Other options for a nuclear spatial profile like the Woods-Saxon model were not considered in this work. The increase of the maximum impact parameter $b_{max}$ allowed for nuclear targets is accounted for by the substitution of $\sigma_{nd}$ in \eq{bmax} by the cross section values for d-Au and p-Pb collisions given in \cite{Kopeliovich:2003tz,Khachatryan:2015zaa}.

Finally, the hadronization of the scattered partons into the finally observed hadrons is described in terms of the Lund string fragmentation model as embedded in the PYTHIA event generator. More specifically, PYTHIA6 \cite{Sjostrand:2006za} is used to arrange partons resulting from hard scattering and initial and final state radiation processes into strings; PYTHIA 8.186 \cite{Sjostrand:2007gs} is used to simulate their fragmentation into hadrons in the framework of the popcorn model. This particular choice of hadronization model turns out to be crucial for a good description of the data. Other possible choices, like the diquark model, result in much softer spectra of the produced pions, yielding a worse agreement with data.
%The results of this method are compared with those obtained using fragmentation functions.
The remnants of the colliding hadrons are also arranged into strings (stretched between quark-diquark pairs). The fraction of the total energy $x$ carried by the quark is chosen according to the probability density:
	\begin{equation}\label{xProb}
		P(x) \propto \frac{(1-x)^{\alpha}}{\sqrt[4]{x^{2}+c^{2}_{min}}}.
	\end{equation}
Where $c_{min}=2\langle m_{q} \rangle/ \sqrt{s}=0.6/ \sqrt{s}$. For the $\alpha$ parameter we use the PYTHIA6 default value $\alpha=3$. No primordial $k_{t}$ distribution is considered, as in \cite{Deng:2014vda} it was shown to be unnecessary in this framework.
%Especificar si cambiamos alpha para LHCf o no

\section{Inclusive hadron transverse momentum spectra at RHIC}\label{RHIC}
\begin{figure}
	\centering
	\includegraphics[width=0.5\textwidth]{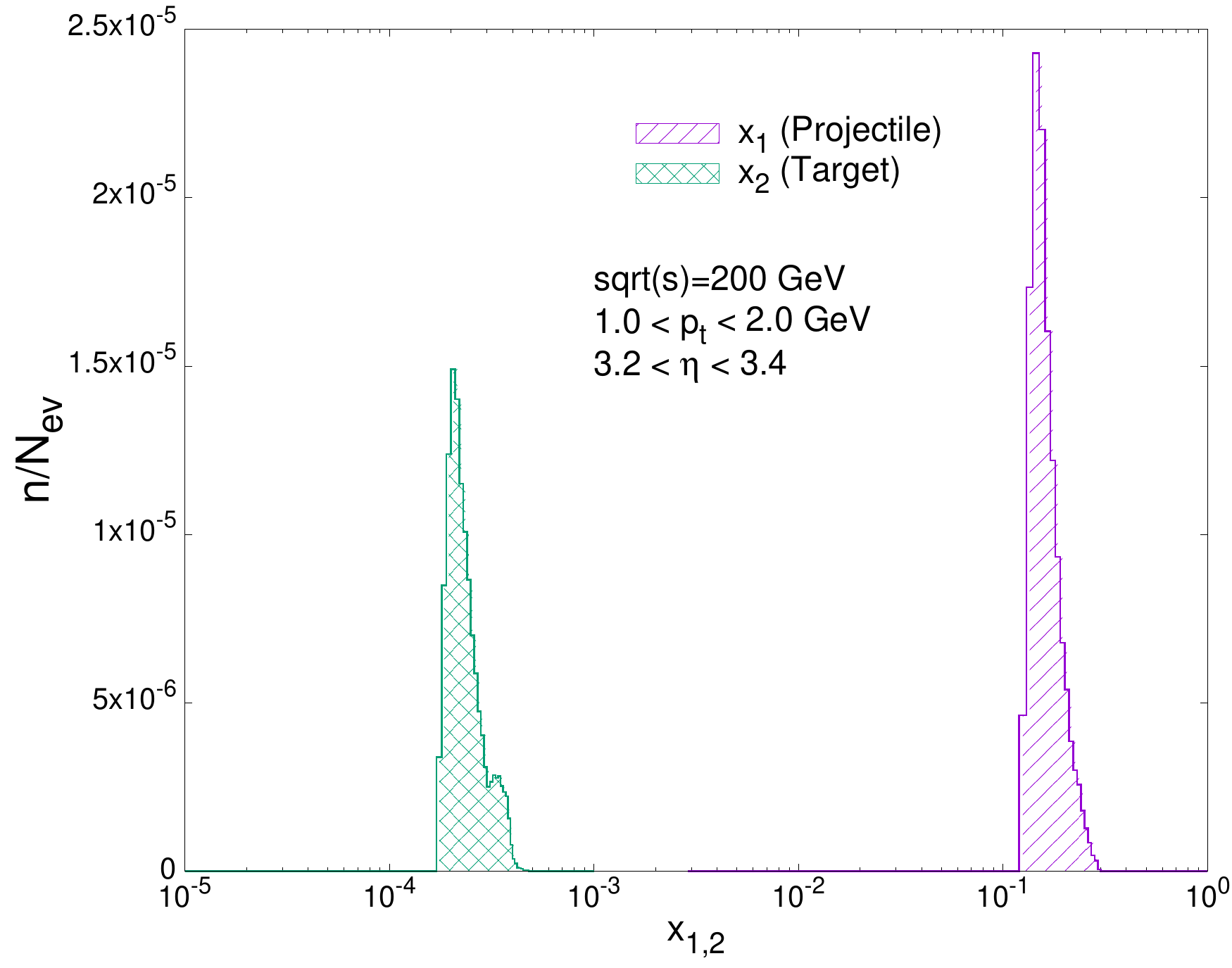}
	\caption{Average distribution of Bjorken-$x$ values for projectile and target and pseudorapidity of the produced particle $3.2\leq y\leq 3.4$.}
	\label{xRHIC}
	\end{figure}

In this section we compare our results to experimental data in the kinematic range observed by two different RHIC detectors, namely BRAHMS \cite{Arsene:2004ux} and STAR \cite{Adams:2006uz}. The kinematic conditions achieved in the deuterium-gold collisions performed at $\sqrt{s}=200$ GeV at RHIC are appropriate for a description in terms of the DHJ formula, provided that we focus on the high-rapidity region of the spectra, as depicted in Fig. \ref{xRHIC}. These fits act as a reference for calibration, as the datasets used have been largely studied in previous works based on the DHJ formula \cite{Albacete:2010bs,Deng:2014vda,JalilianMarian:2011dt}. We build the PDF of the deuteron from the proton PDFs assuming strict isospin symmetry.
For each independent hard scattering we calculate the multiplicity density of produced particles from the \eq{DHJ} for the cross section scaling by the non-diffractive cross-section:
\begin{equation}\label{multi}
\frac{dN^{h_1h_2\to (q/g)X}}{dy\,d^2k}=\frac{1}{\sigma_{nd}}\frac{d\sigma^{h_1h_2\to (q/g)X}}{dy\,d^2k}.
\end{equation}
 
We assume that the energy dependence of the non-diffractive cross section in \eq{multi} cancels off the energy dependence of the $\sigma_0$ factor in \eq{DHJ}, even if these two objects are not necessarily the same one. Any possible deviation from this assumption is lumped on the corresponding $K$-factors. 
 
%For the sake of completeness we compare the Monte-Carlo output with results obtained from a different approach for hadronization: the use of fragmentation functions.

%	\begin{figure}
%	\centering
%	\begin{subfigure}{0.48\textwidth}
%	\centering
%	\includegraphics[width=1.05\textwidth]{Figs/2rhicpp_70M_events/Brahms_Phenix_pp.pdf}
%	\end{subfigure}
%	\subfigure{\label{fig:1a}\includegraphics[width=0.48\textwidth]{Figs/2rhicpp_70M_events/Brahms_Phenix_pp.pdf}}
%	\subfigure{\label{fig:1b}\includegraphics[width=0.48\textwidth]{Figs/2rhicpp_70M_events/Star_pp.pdf}}	
%	\begin{subfigure}{0.48\textwidth}
%	\centering
%	\includegraphics[width=1.05\textwidth]{Figs/2rhicpp_70M_events/Star_pp.pdf}
%	\end{subfigure}
%	\caption{Left plot: negatively charged hadron transverse momentum spectra at pseudorapidities $\eta=$2.2 and 3.2 (measured at BRAHMS detector) and neutral pion spectra in the pseudorapidity range $3.2<\eta<3.8$ (measured at PHENIX detector) in p-p collisions at $\sqrt{s}=200\, \text{GeV}$. Right plot: neutral pion spectra at pseudorapidities $\eta=$3.3, 3.8 and 4 in p-p collisions at $\sqrt{s}=200\, \text{GeV}$ measured at STAR detector.}
%	\label{ppRHIC}
%	\end{figure}
	
	\begin{figure}
	\centering
%	\begin{subfigure}{0.48\textwidth}
%	\centering
	\subfigure{\label{fig:2a}\includegraphics[width=0.48\textwidth]{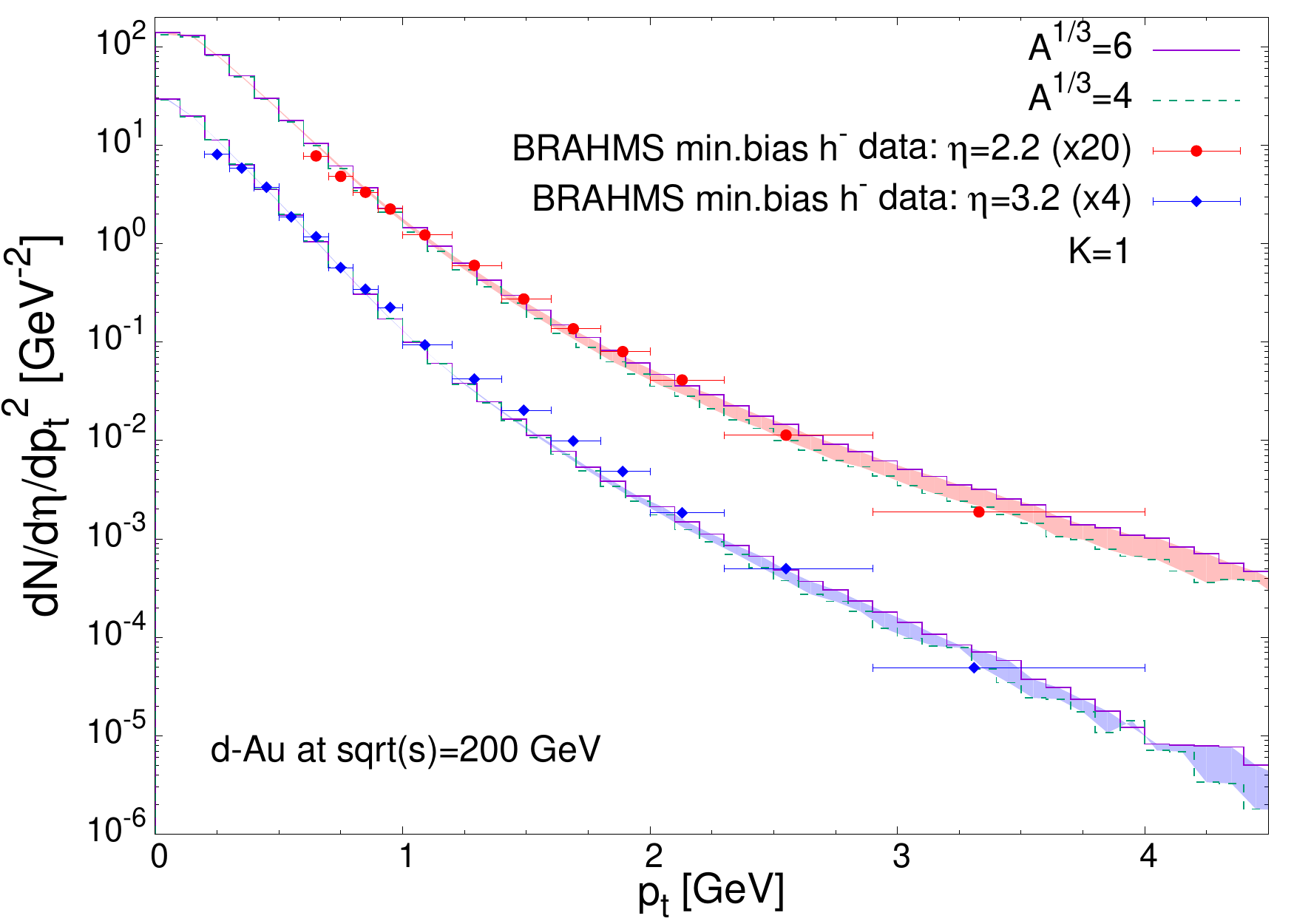}}
	\subfigure{\label{fig:2a}\includegraphics[width=0.48\textwidth]{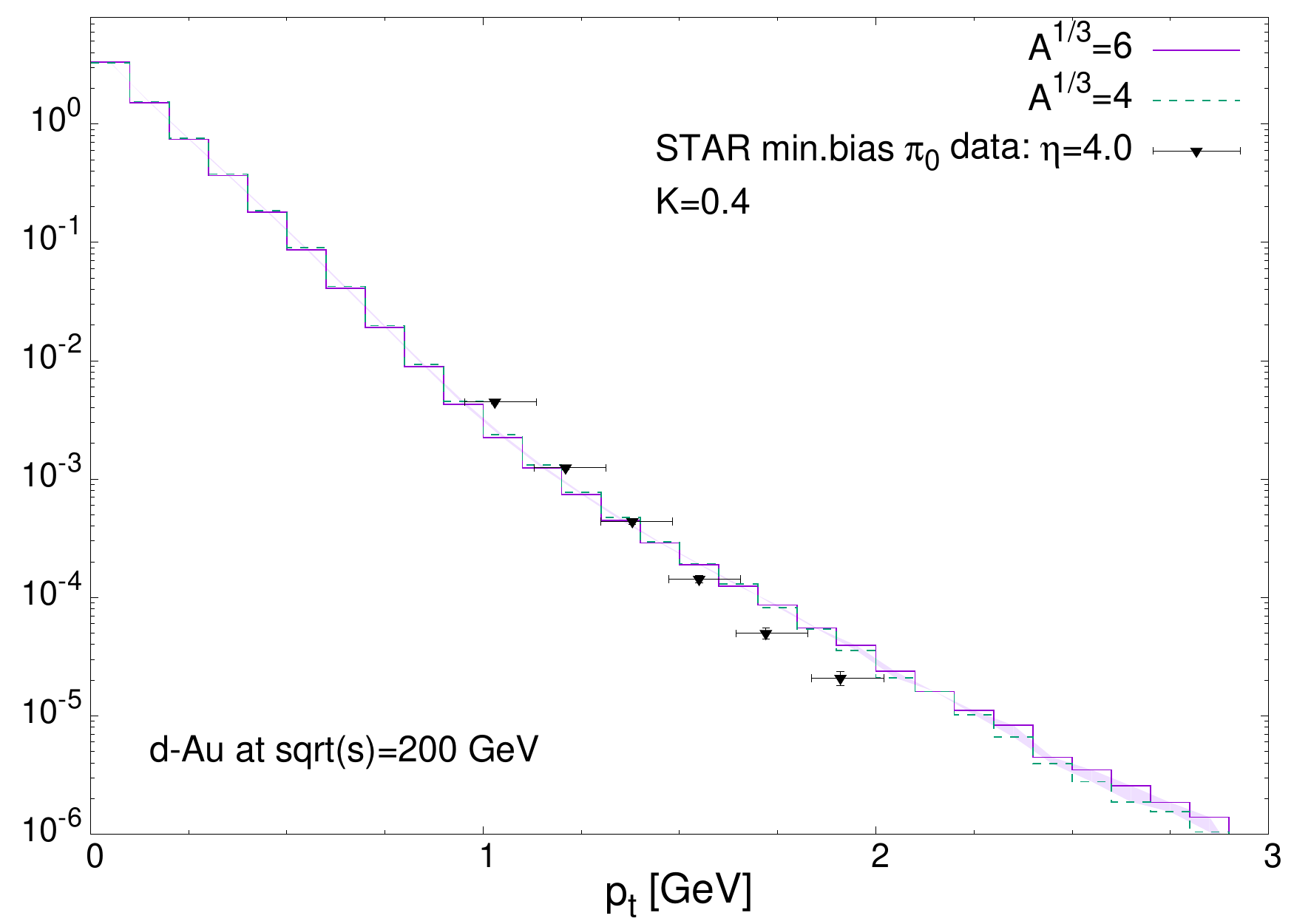}}
%	\end{subfigure}
%	\begin{subfigure}{0.48\textwidth}
%	\centering
%	\includegraphics[width=1.05\textwidth]{Figs/3rhicdAU_20M_events/Star_dAu.pdf}
%	\end{subfigure}
	\caption{Left plot: negatively charged hadron transverse momentum spectra at  $\eta=$2.2 and 3.2 in d-Au collisions at $\sqrt{s}=200\, \text{GeV}$ measured by the BRAHMS collaboration. Right plot: neutral pion spectra at $\eta=$4 in d-Au collisions at $\sqrt{s}=200\, \text{GeV}$ measured by the STAR collaboration. Scale dependence between $Q_{s0}^{2}=0.157\,A^{1/3}\,\text{GeV}^{2}$ with $A^{1/3}=6$ and 4 is shown by the shaded areas.
%	%Scale dependence between $\mu=\text{max}\{ k_{t}, Q_{s} \}$, $2\mu$ and $\mu/2$ is almost negligible in the whole momentum range.
%	%En este plot o en el siguiente: dependence with $\gamma$.
	}
	\label{dAURHIC}
	\end{figure}

We reach a rather good description of d-Au data on the spectra of negatively charged hadrons measured at pseudorapidities $\eta\!=\!2.2$ and 3.2 by BRAHMS in minimum bias collisions and also of STAR data on neutral pion production at $\eta\!=\!4$, see Fig \ref{dAURHIC}. Our results are little sensitive to the specific value of the number of participants nucleons in the collision which, in the mean field treatment of nuclear geometry performed here, is given by $N_{part}\!\approx\!A^{1/3}$. The most remarkable, and completely new, feature of our result is that, by means of the Lund fragmentation mechanism implemented in our Monte Carlo, we can reach values of the transverse momentum of the produced particle as low as detected experimentally $p_{t,min}\sim 0.2$ GeV. Previous approaches relied on the use of fragmentation functions to describe the hadronization process. Hence, by construction, they could only access the regime of perturbatively high transverse momenta $p_{t,min}\gtrsim 1$ GeV, where these functions are defined. BRAHMS data is well described with a $K$-factor $K=1$. However, STAR data on neutral pions can only be described with a $K$-factor $K=0.4$, exactly the same value obtained in previous analysis of data.

%In order to reproduce STAR data we need an overall normalization of $K=0.25$. On the other hand, data measured at BRAHMS and PHENIX are well reproduced using $K=1$. The physical meaning of the overall $K$-factors can be understood as the contribution of two different effects not included in this approach; one of geometric nature, and other one due to us using the leading-order (LO) expression of the DHJ formula and thus neglecting higher twist corrections.
%	\begin{equation}\label{Kfactor}
%		K=K_{geometric}\,K_{dynamical}
%	\end{equation}
%$K_{geometric}$ accounts for a non theoretically determined yet dependence of the average transverse area of the proton ($\sigma_{0}/2$) with $\sqrt{s}$. In this approach we use the result obtained in \cite{Albacete:2009fh} by fits to data from deep inelastic scattering experiments at HERA ($\sigma_{0}/2=16.5\,\text{mb}$). Regarding $K_{dynamical}$, the next-to-leading order (NLO) corrections obtained in \cite{Stasto:2014sea} were found to yield a similar dependence in $p_{T}$ as the LO expression for a gluon saturated regime, making it reasonable to approximate their effect with a constant $K$-factor. In this paper we do not quantitatively factorise $K$ into these two contributions, but rather raise the possibility of a $K=1$ value being due to the non-trivial cancellation between them.

\section{Inclusive hadron transverse momentum spectra at LHCf}
\begin{figure}[htbp]
	\centering
	\includegraphics[width=0.5\textwidth]{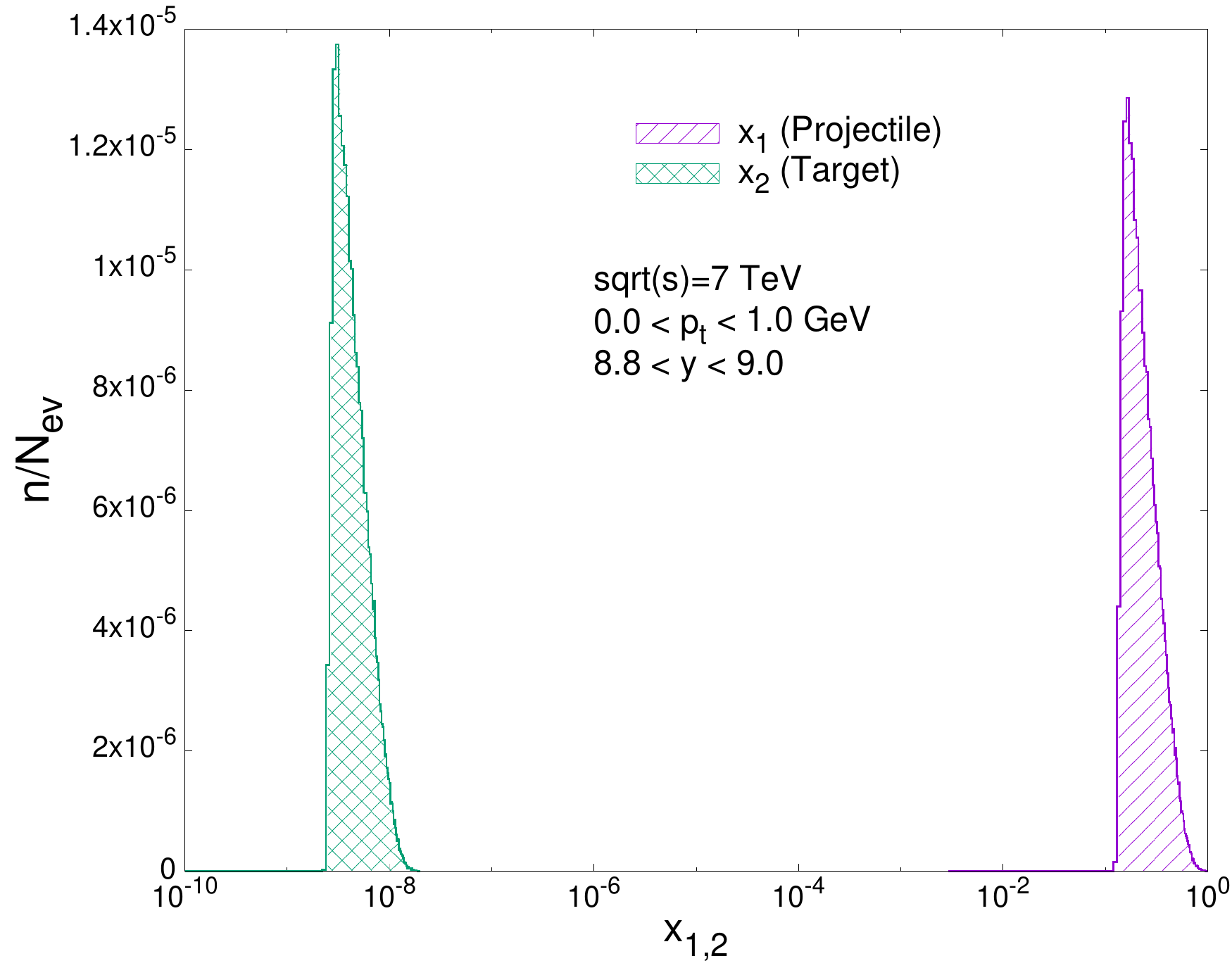}
	\caption{Average distribution of Bjorken-$x$ values for projectile and target and rapidity of the produced particle $8.8\leq y\leq 9.0$.}
	\label{xLHCf}
	\end{figure}

In this section we compare our results with the preliminary data on neutral pion production measured by the LHCf collaboration in p-p and p-Pb collisions at $\sqrt{s}=7\,\text{TeV}$ and $5.02\,\text{TeV}$ respectively \cite{Adriani:2015iwv}. The ultra-high rapidity range available in this experiment ($8.8\!\leq\! y\!\leq\! 10.8$) is appropriate for a description in terms of the DHJ formula, as shown in Fig. \ref{xLHCf} where we plot the distribution of Bjorken-$x$ values contributing to these collisions from the projectile and target respectively. They are peaked in $x_{proj}\approx0.1$ and $x_{targ}\approx10^{-8}$, which indicates a much stronger dilute-dense asymmetry than in the RHIC case, Fig \ref{xRHIC}.
	
	\begin{figure}[htbp]
	\centering
%	\begin{subfigure}{0.3\textwidth}
%	\centering
		\subfigure{\label{fig:4a}\includegraphics[width=0.3\textwidth]{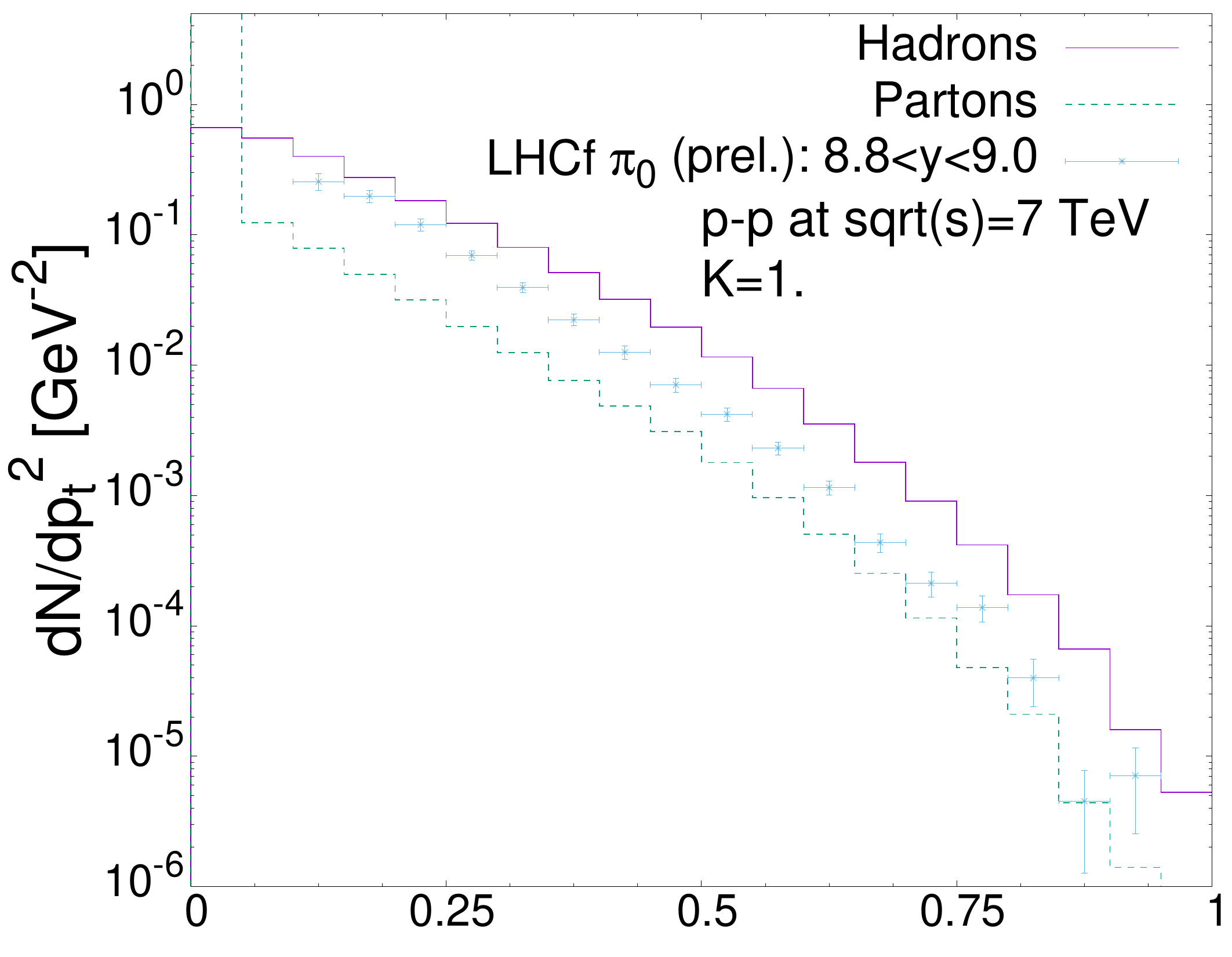}}
		\subfigure{\label{fig:4b}\includegraphics[width=0.3\textwidth]{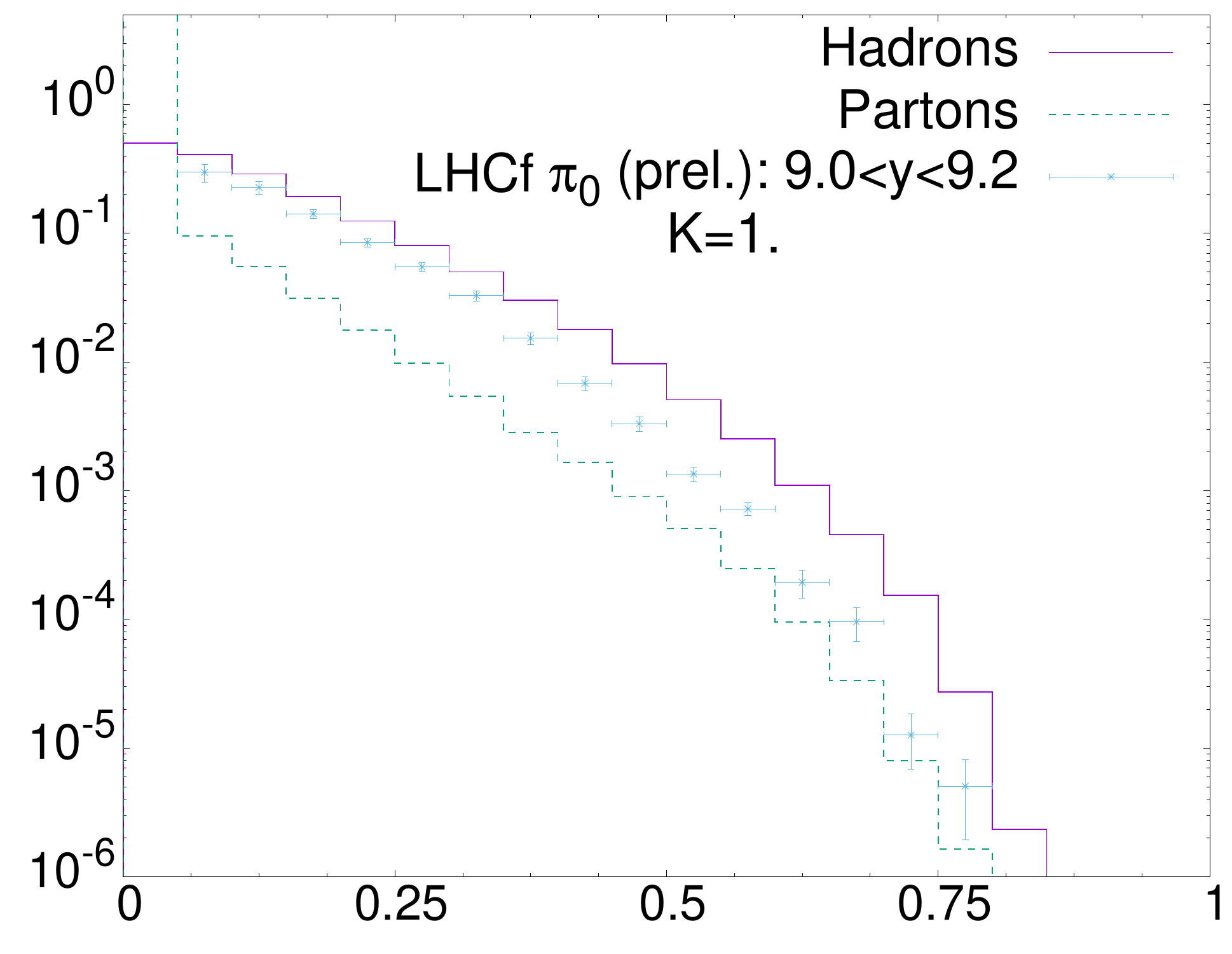}}
		\subfigure{\label{fig:4c}\includegraphics[width=0.3\textwidth]{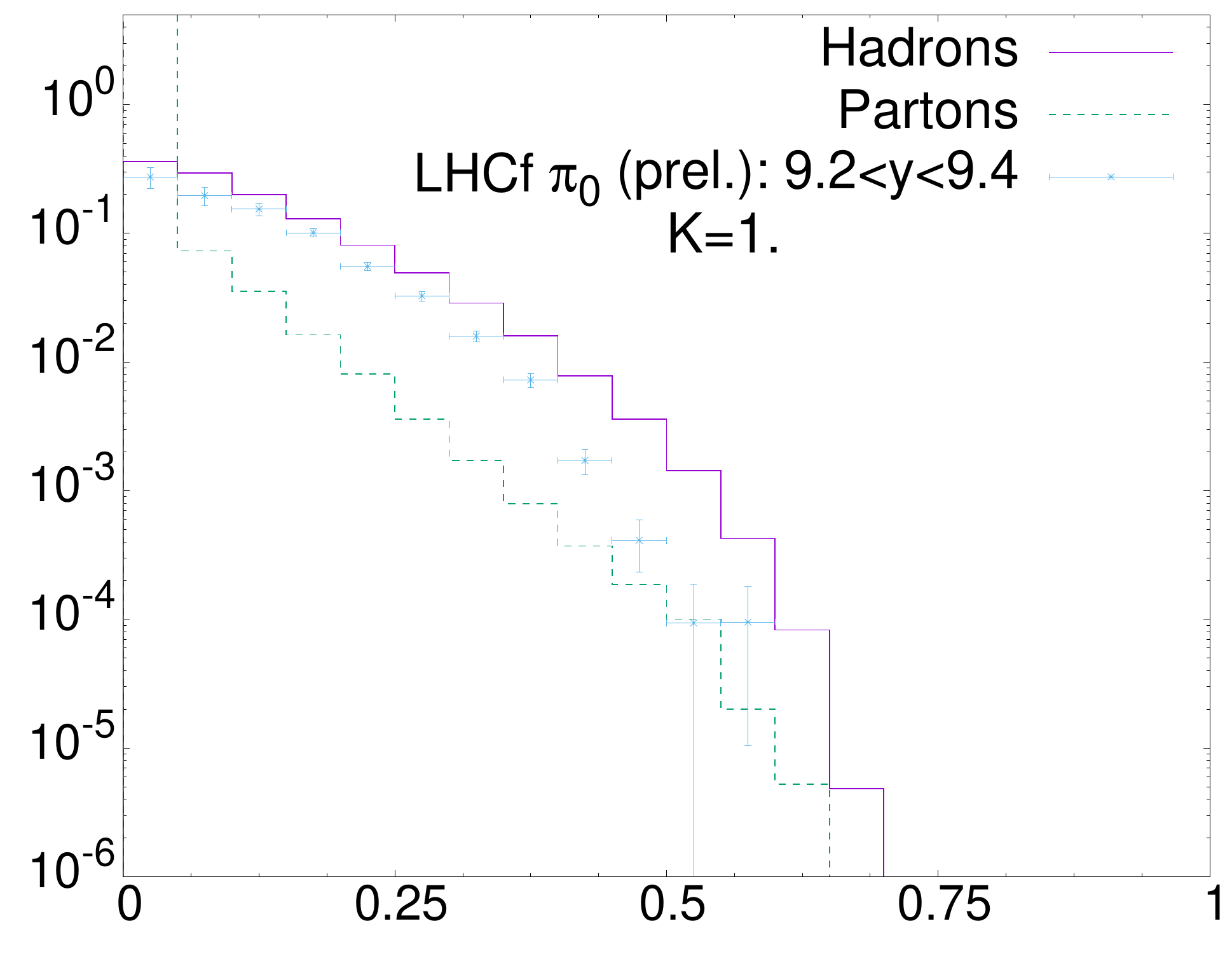}}
		\subfigure{\label{fig:4d}\includegraphics[width=0.3\textwidth]{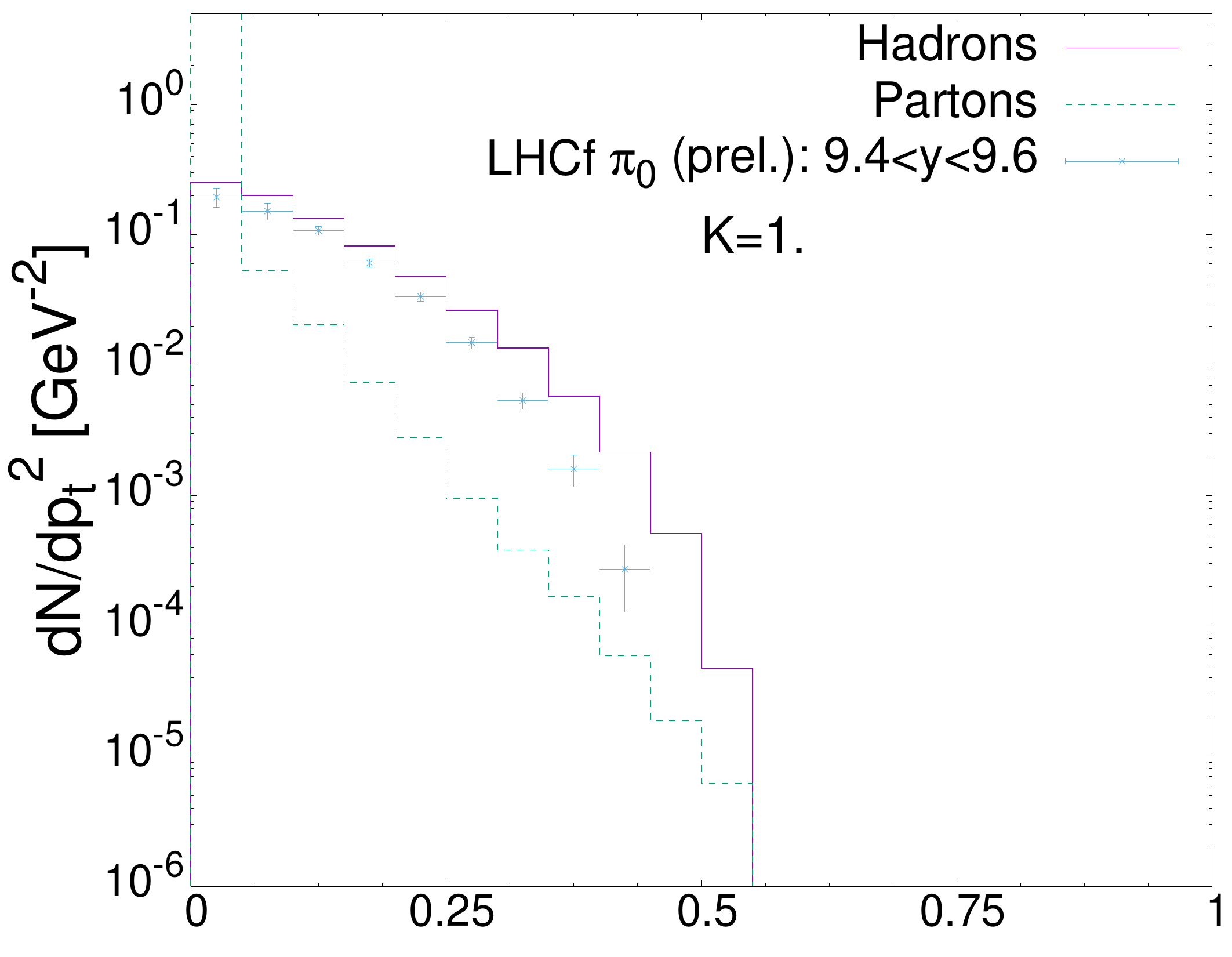}}
		\subfigure{\label{fig:4e}\includegraphics[width=0.3\textwidth]{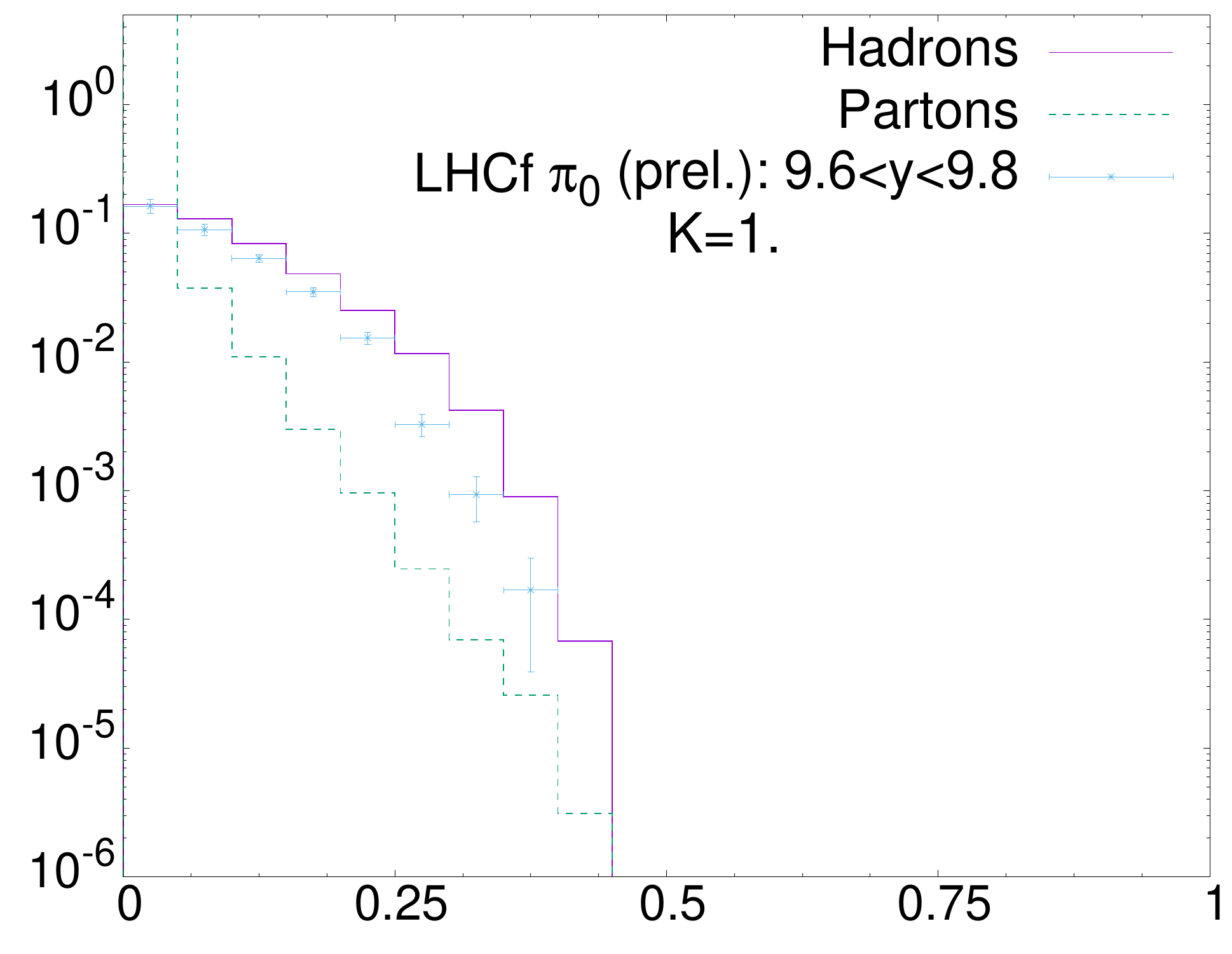}}
		\subfigure{\label{fig:4f}\includegraphics[width=0.3\textwidth]{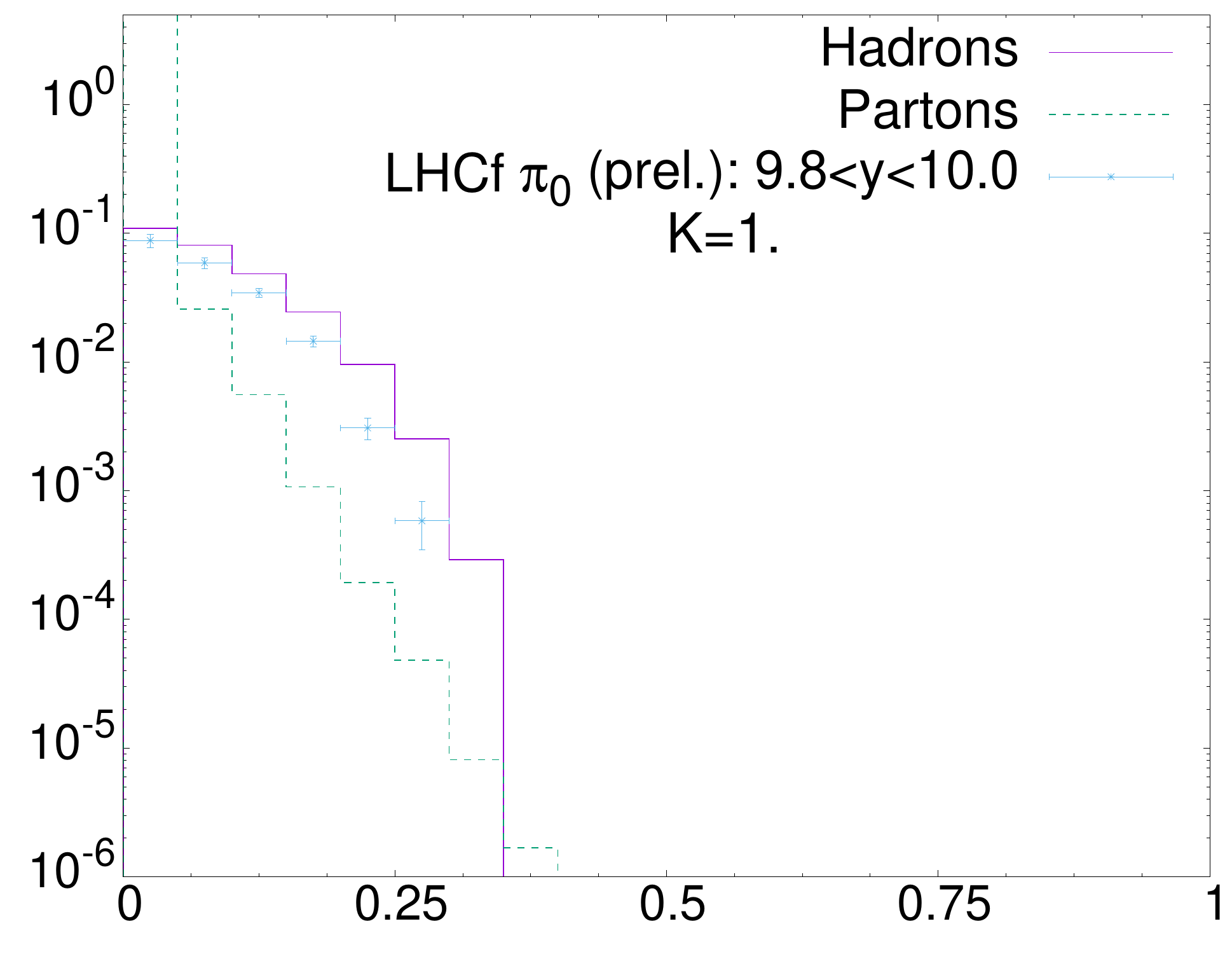}}
		\subfigure{\label{fig:4g}\includegraphics[width=0.3\textwidth]{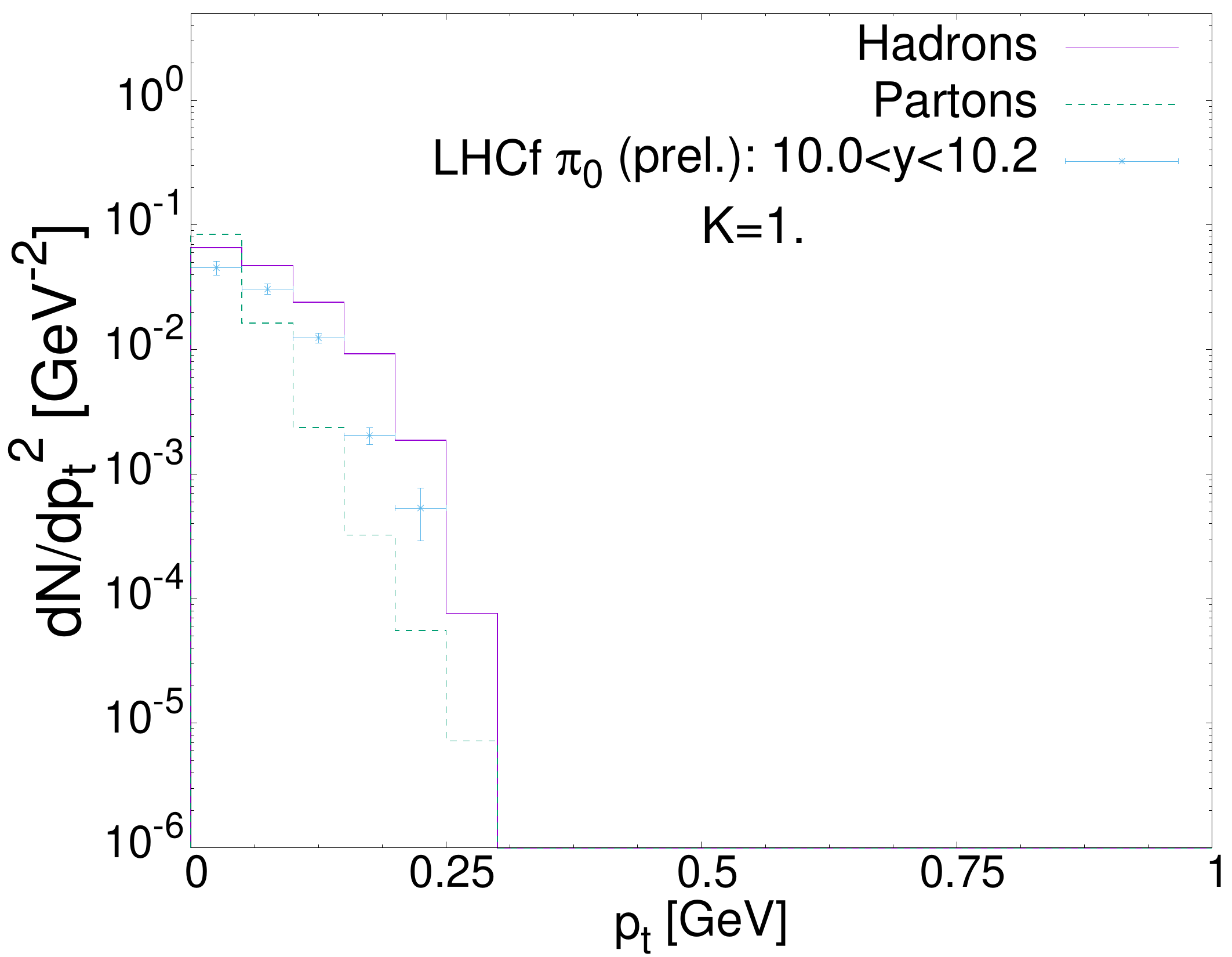}}
		\subfigure{\label{fig:4h}\includegraphics[width=0.3\textwidth]{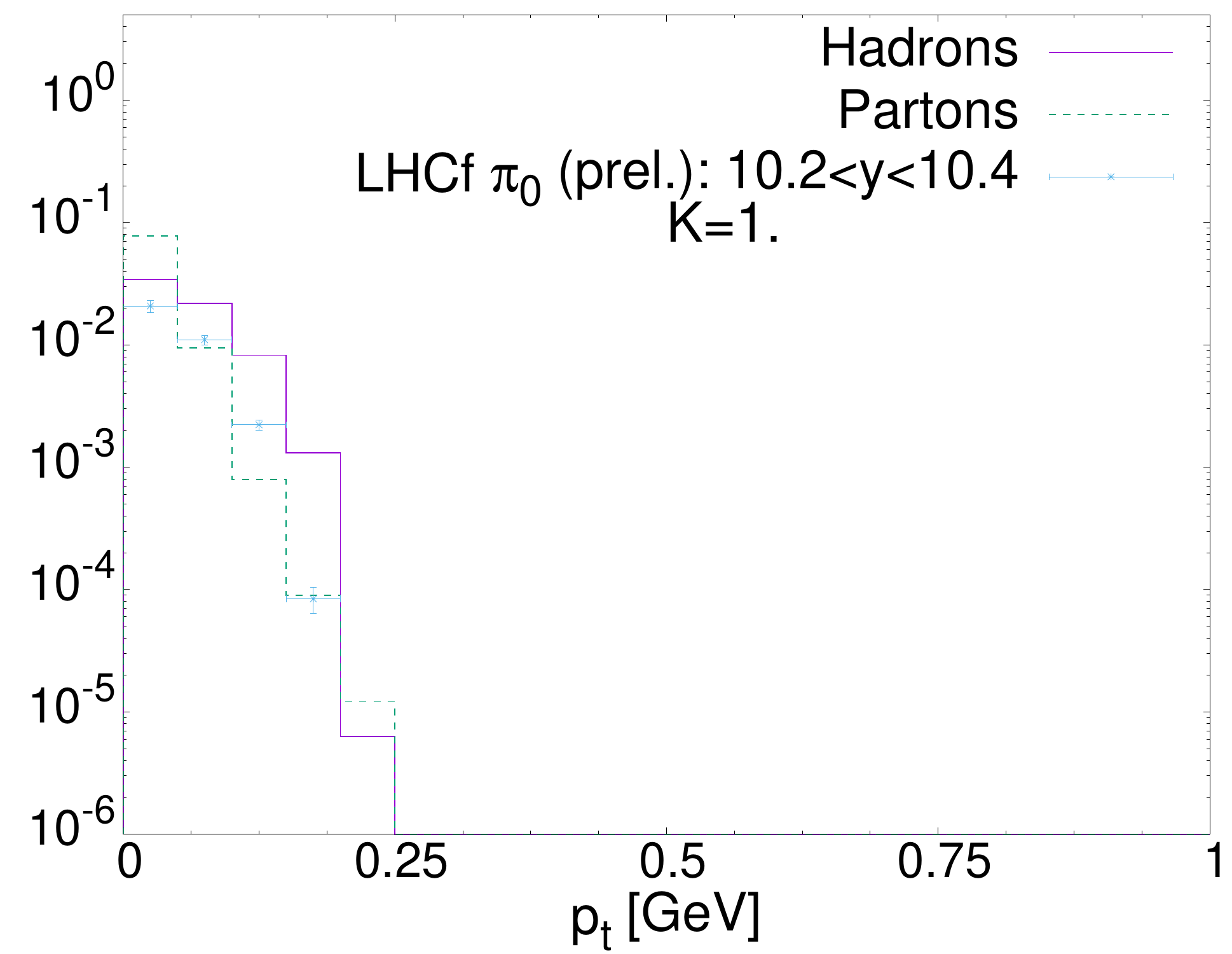}}
		\subfigure{\label{fig:4i}\includegraphics[width=0.3\textwidth]{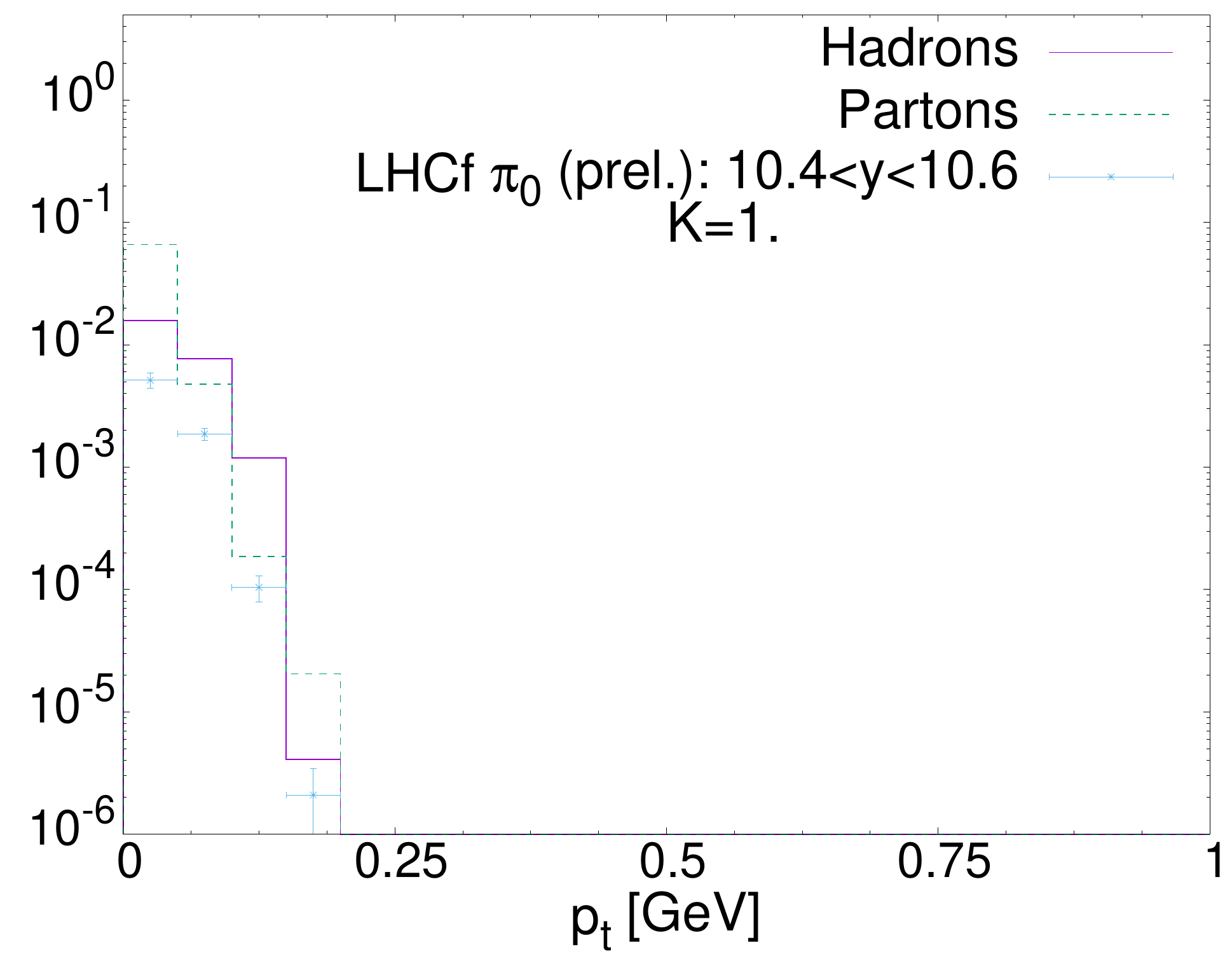}}		
%	\includegraphics[width=\textwidth]{Figs/7lhcfpp_25M_events/{lhcf_p_p_8.8-9.0}.pdf}
%	\end{subfigure}
%	\begin{subfigure}{0.3\textwidth}
%	\centering
%	\includegraphics[width=\textwidth]{Figs/7lhcfpp_25M_events/{lhcf_p_p_9.0-9.2}.pdf}
%	\end{subfigure}
%	\begin{subfigure}{0.3\textwidth}
%	\centering
%	\includegraphics[width=\textwidth]{Figs/7lhcfpp_25M_events/{lhcf_p_p_9.2-9.4}.pdf}
%	\end{subfigure}
%	\begin{subfigure}{0.3\textwidth}
%	\centering
%	\includegraphics[width=\textwidth]{Figs/7lhcfpp_25M_events/{lhcf_p_p_9.4-9.6}.pdf}
%	\end{subfigure}
%	\begin{subfigure}{0.3\textwidth}
%	\centering
%	\includegraphics[width=\textwidth]{Figs/7lhcfpp_25M_events/{lhcf_p_p_9.6-9.8}.pdf}
%	\end{subfigure}
%	\begin{subfigure}{0.3\textwidth}
%	\centering
%	\includegraphics[width=\textwidth]{Figs/7lhcfpp_25M_events/{lhcf_p_p_9.8-10.0}.pdf}
%	\end{subfigure}
%	\begin{subfigure}{0.3\textwidth}
%	\centering
%	\includegraphics[width=\textwidth]{Figs/7lhcfpp_25M_events/{lhcf_p_p_10.0-10.2}.pdf}
%	\end{subfigure}
%	\begin{subfigure}{0.3\textwidth}
%	\centering
%	\includegraphics[width=\textwidth]{Figs/7lhcfpp_25M_events/{lhcf_p_p_10.2-10.4}.pdf}
%	\end{subfigure}
%	\begin{subfigure}{0.3\textwidth}
%	\centering
%	\includegraphics[width=\textwidth]{Figs/7lhcfpp_25M_events/{lhcf_p_p_10.4-10.6}.pdf}
%	\end{subfigure}
	\caption{Neutral pion transverse momentum spectra in the rapidity range $8.9<y<10.6$ in p-p collisions at $\sqrt{s}=7\,\text{TeV}$. Also shown is the corresponding partonic spectra (dashed lines).}
	\label{ppLHCf}
	\end{figure}

	\begin{figure}[htbp]
	\centering
	
		\subfigure{\label{fig:5a}\includegraphics[width=0.3\textwidth]{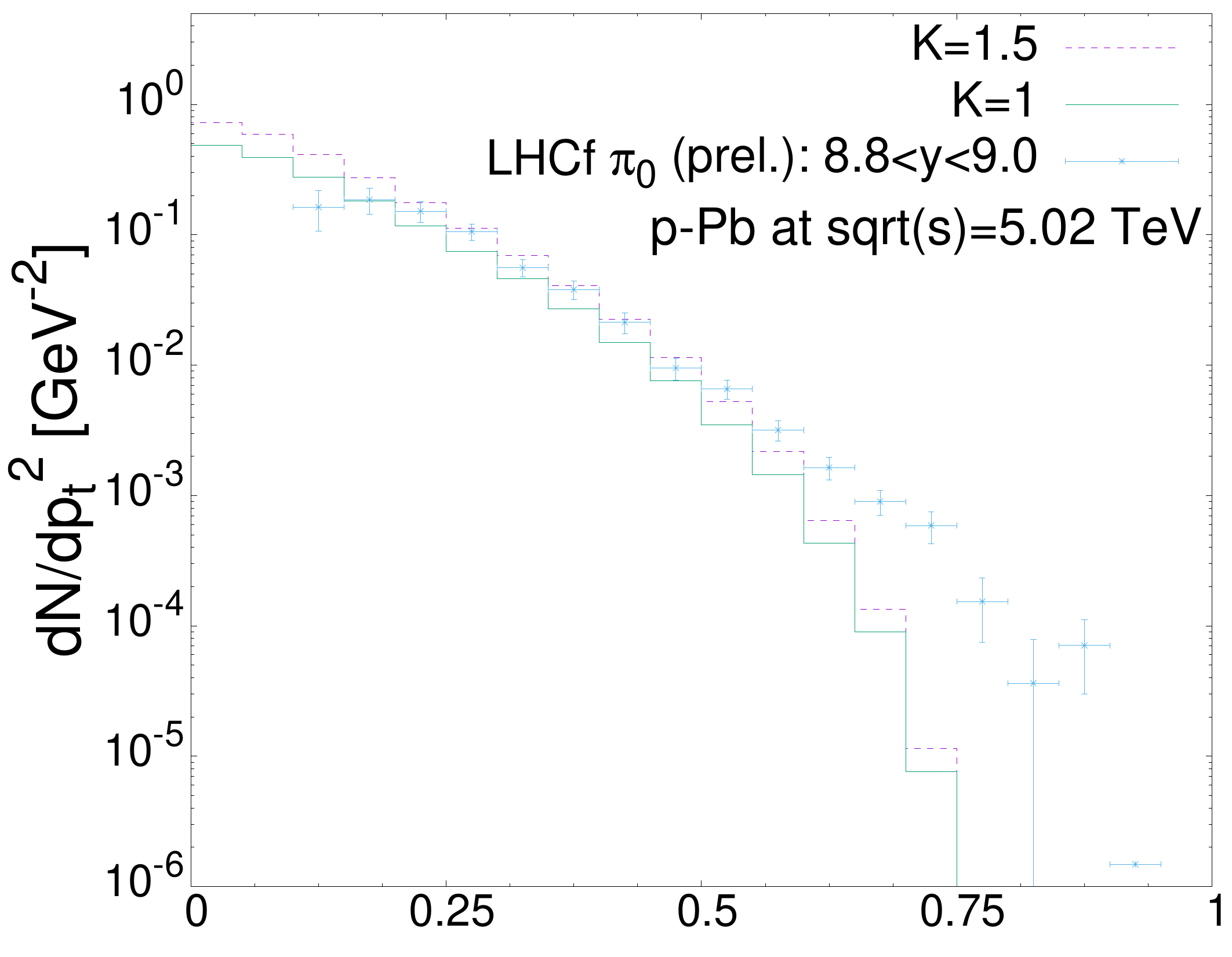}}
		\subfigure{\label{fig:5b}\includegraphics[width=0.3\textwidth]{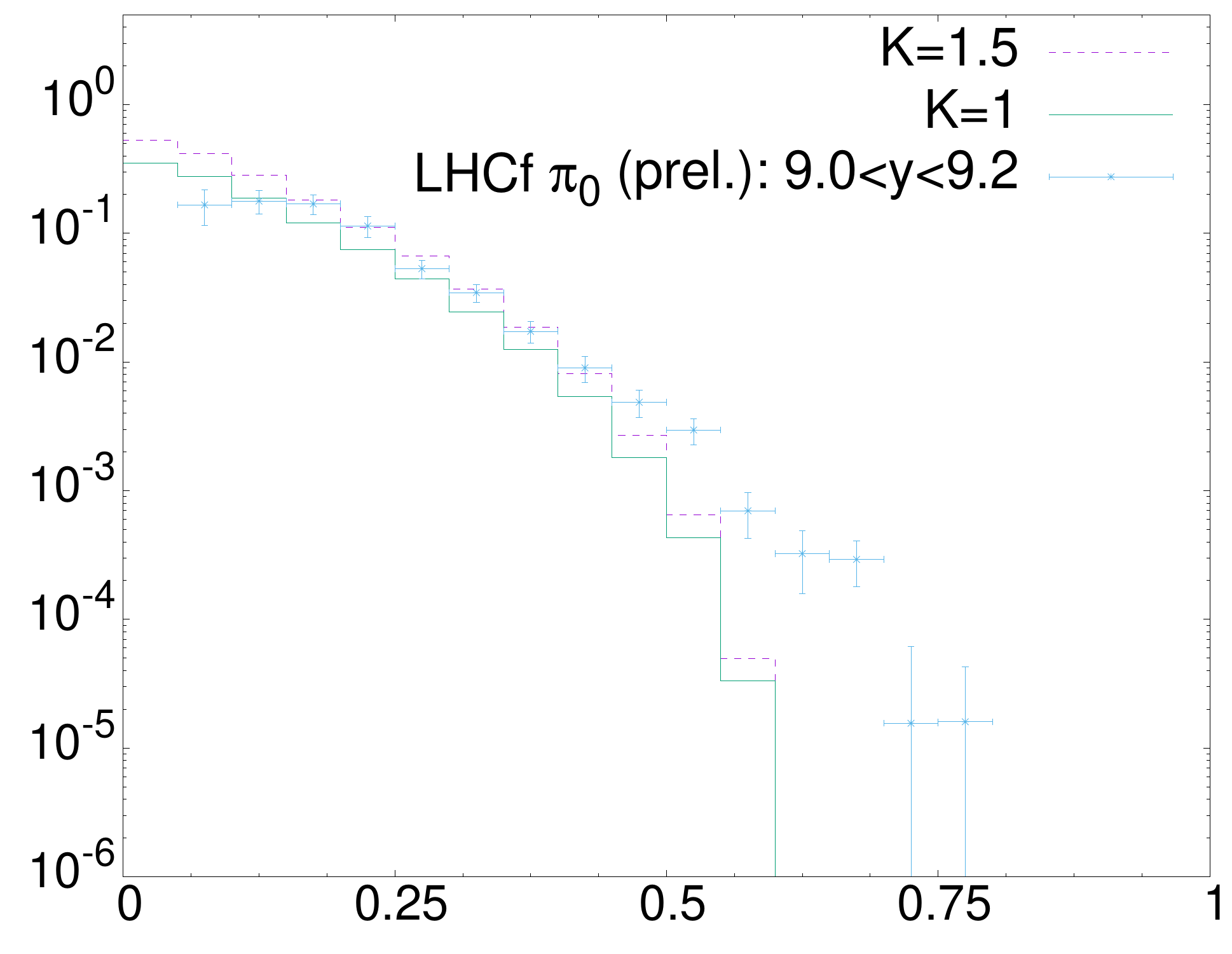}}
		\subfigure{\label{fig:5c}\includegraphics[width=0.3\textwidth]{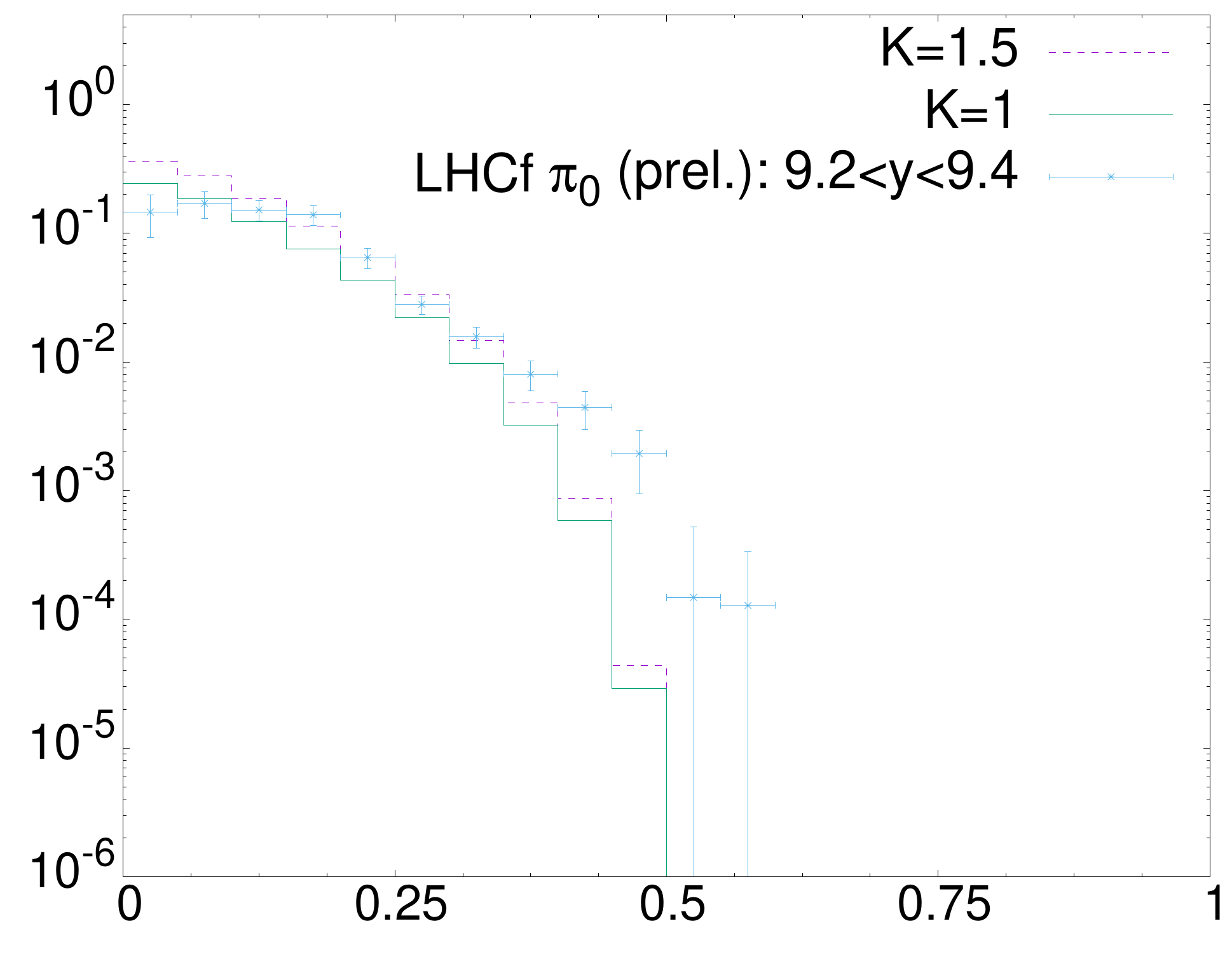}}
		\subfigure{\label{fig:5d}\includegraphics[width=0.3\textwidth]{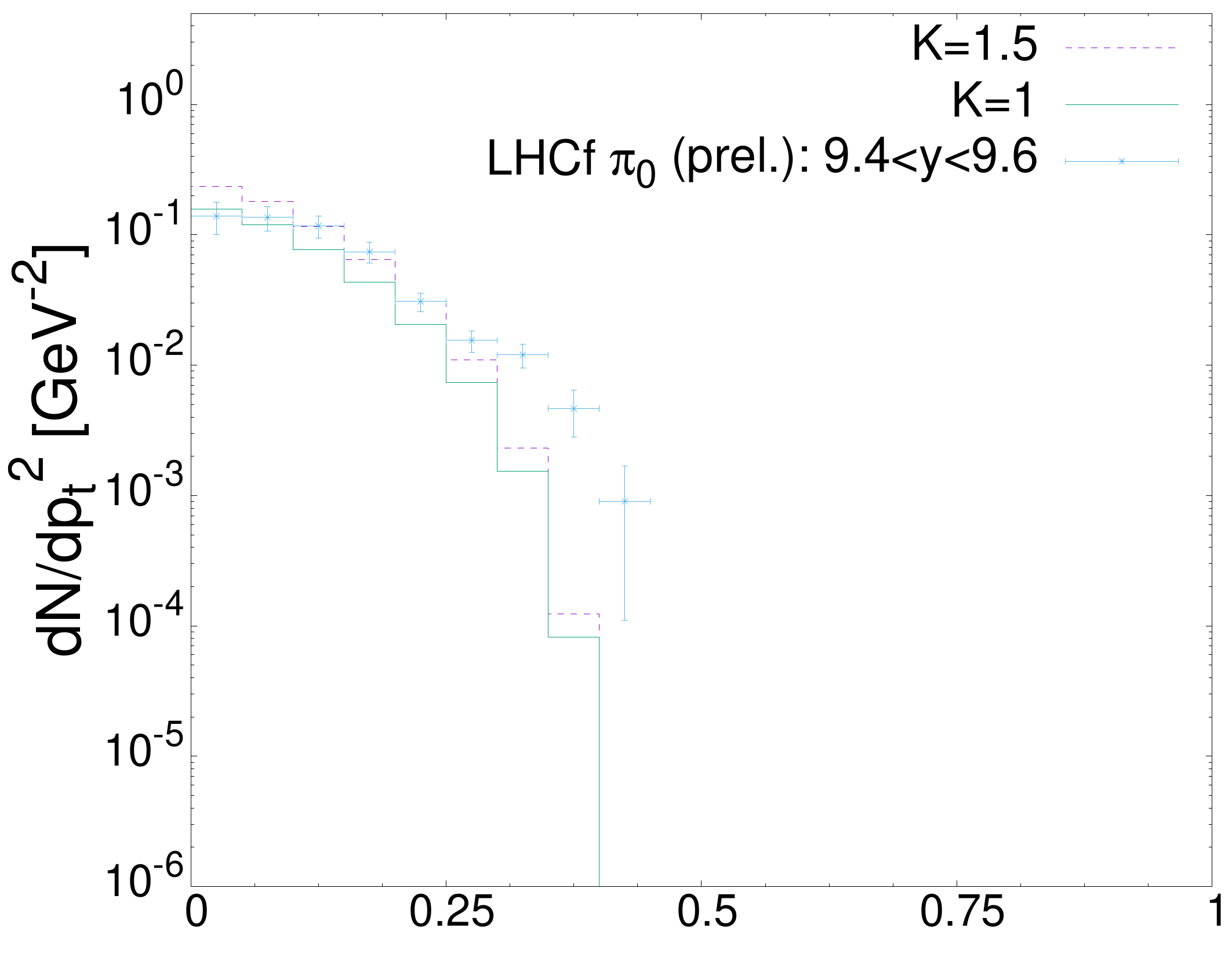}}
		\subfigure{\label{fig:5e}\includegraphics[width=0.3\textwidth]{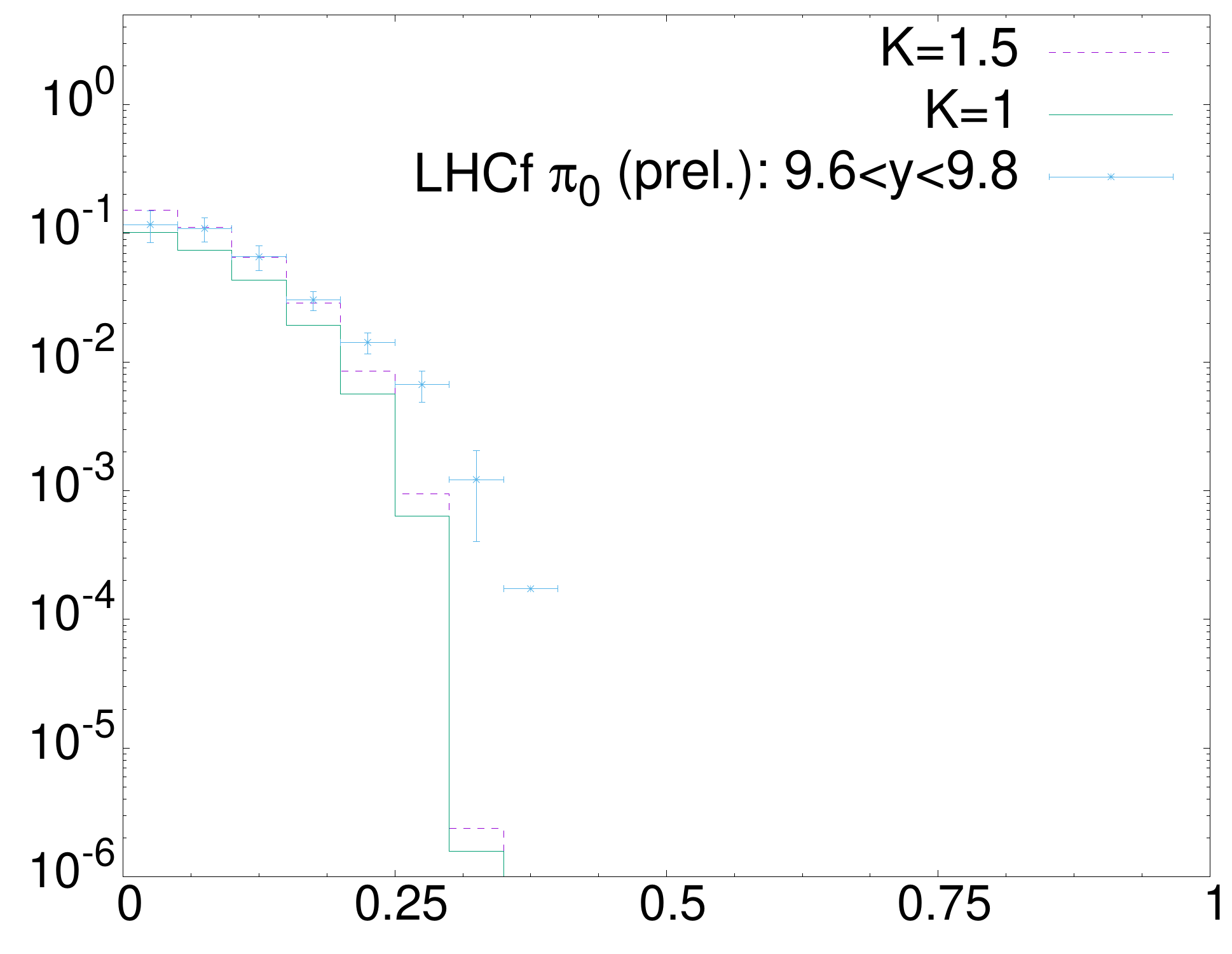}}
		\subfigure{\label{fig:5f}\includegraphics[width=0.3\textwidth]{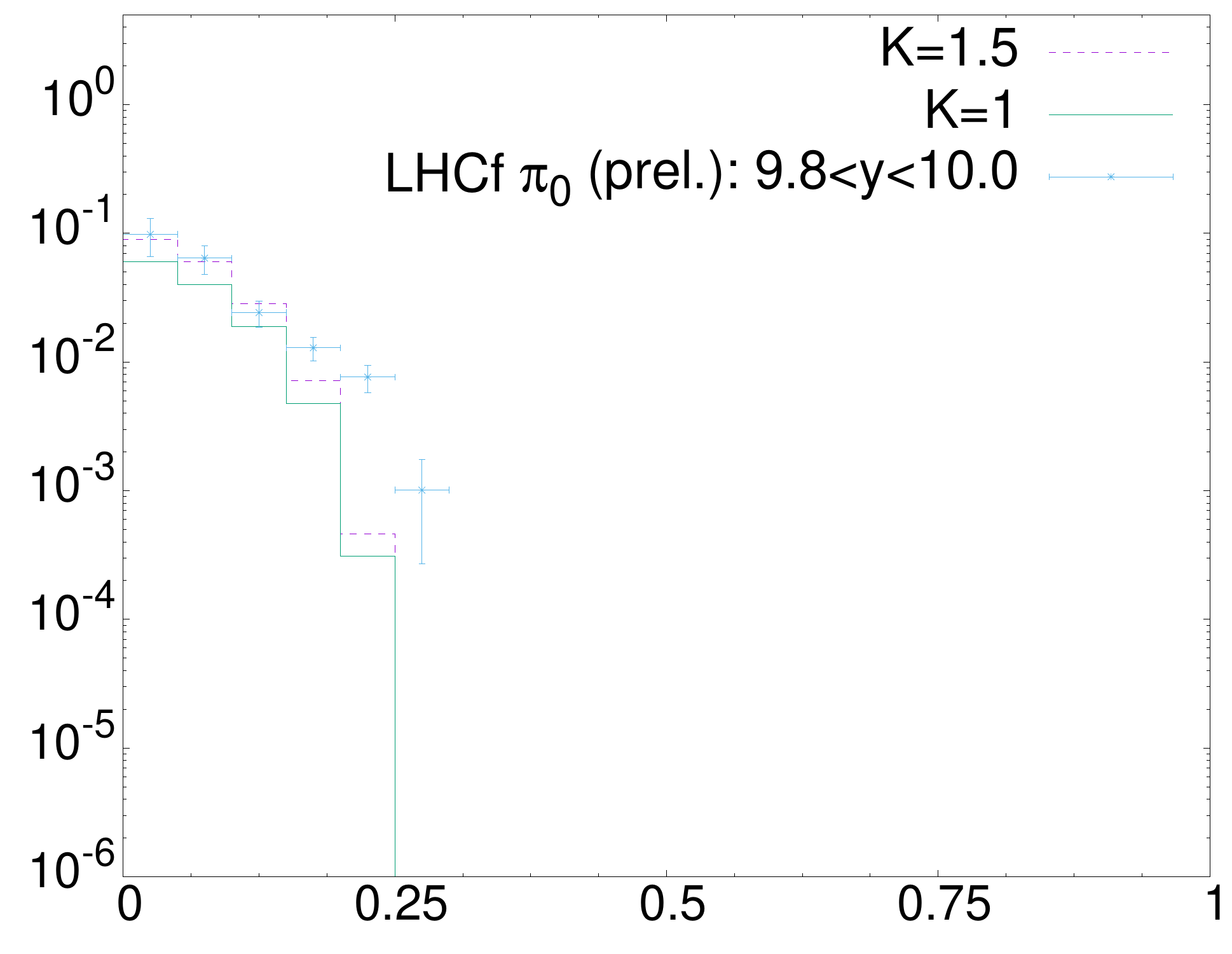}}
		\subfigure{\label{fig:5g}\includegraphics[width=0.3\textwidth]{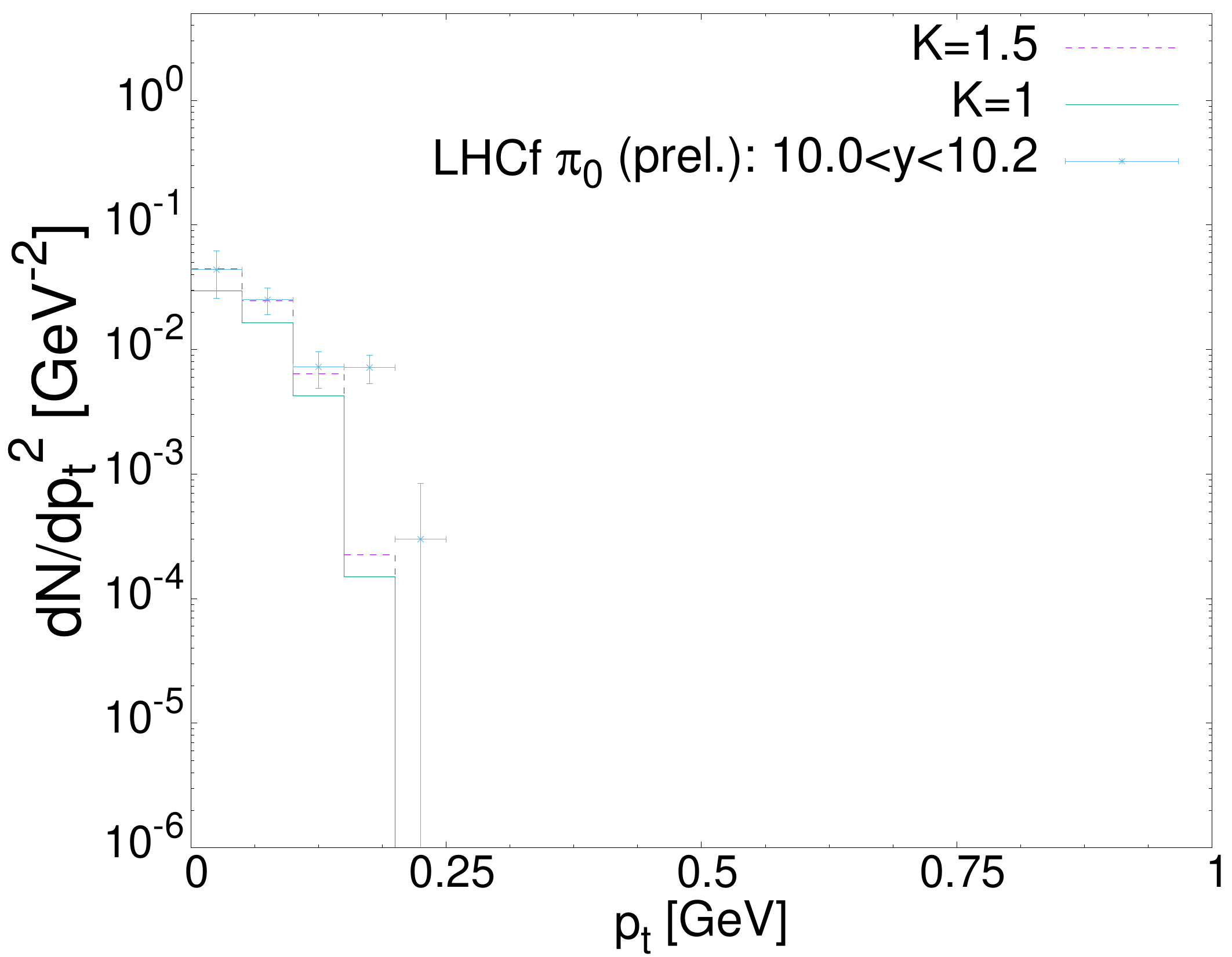}}
		\subfigure{\label{fig:5h}\includegraphics[width=0.3\textwidth]{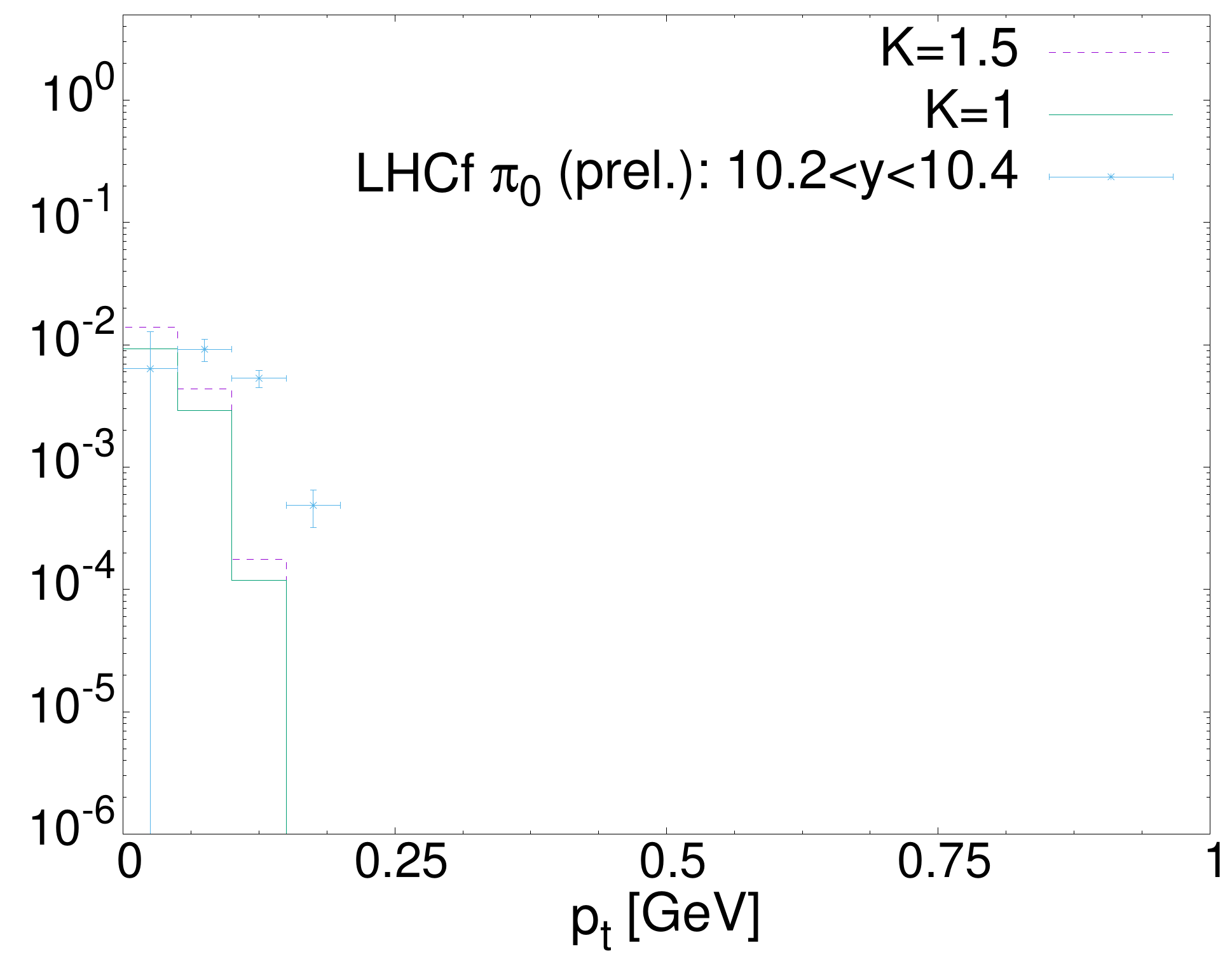}}
		\subfigure{\label{fig:5i}\includegraphics[width=0.3\textwidth]{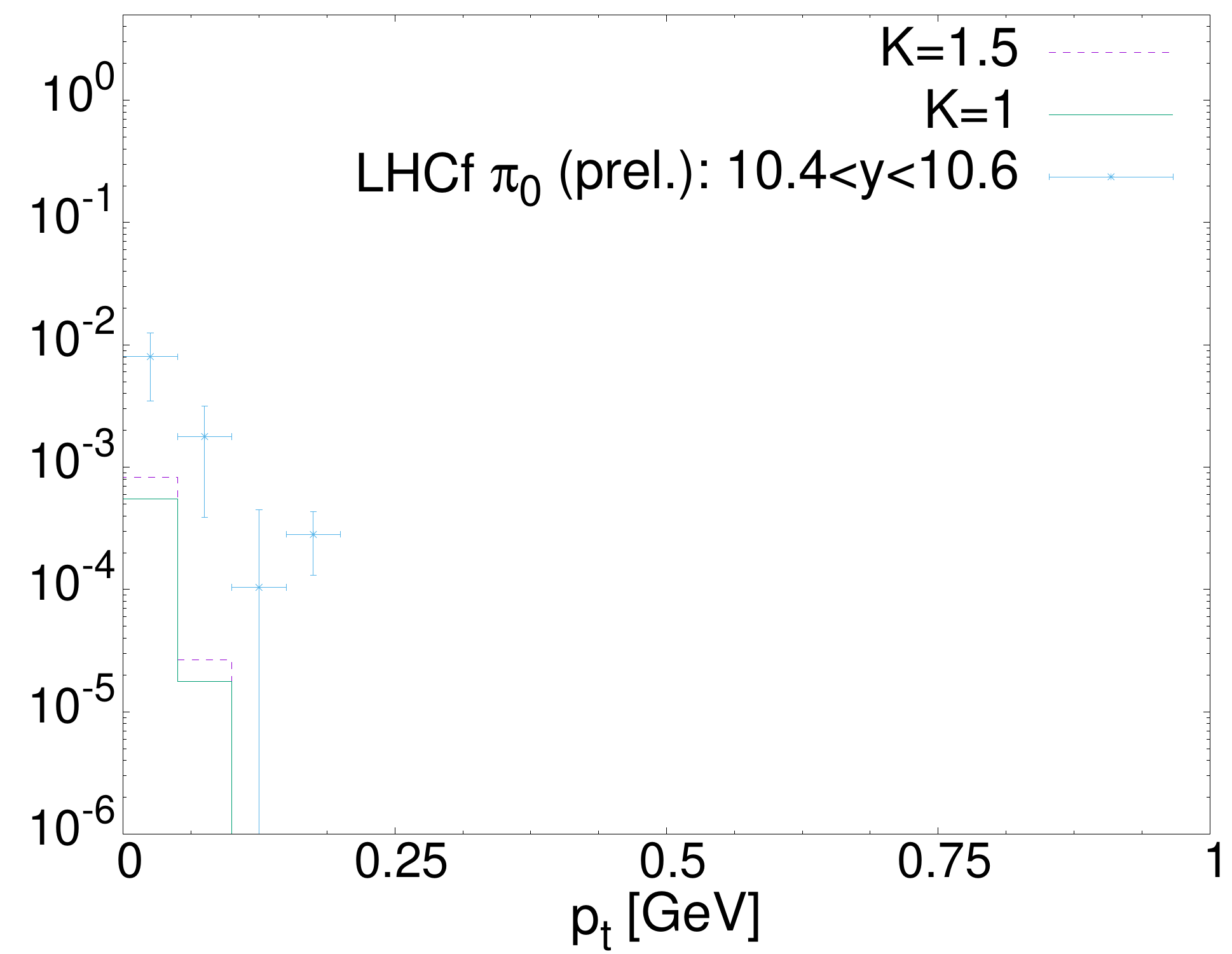}}		
%	\begin{subfigure}{0.3\textwidth}
%	\centering
%	\includegraphics[width=\textwidth]{Figs/4lhcfpPB_100M_events/{lhcf_P_pB_8.8-9.0}.pdf}
%	\end{subfigure}
%	\begin{subfigure}{0.3\textwidth}
%	\centering
%	\includegraphics[width=\textwidth]{Figs/4lhcfpPB_100M_events/{lhcf_P_pB_9.0-9.2}.pdf}
%	\end{subfigure}
%	\begin{subfigure}{0.3\textwidth}
%	\centering
%	\includegraphics[width=\textwidth]{Figs/4lhcfpPB_100M_events/{lhcf_P_pB_9.2-9.4}.pdf}
%	\end{subfigure}
%	\begin{subfigure}{0.3\textwidth}
%	\centering
%	\includegraphics[width=\textwidth]{Figs/4lhcfpPB_100M_events/{lhcf_P_pB_9.4-9.6}.pdf}
%	\end{subfigure}
%	\begin{subfigure}{0.3\textwidth}
%	\centering
%	\includegraphics[width=\textwidth]{Figs/4lhcfpPB_100M_events/{lhcf_P_pB_9.6-9.8}.pdf}
%	\end{subfigure}
%	\begin{subfigure}{0.3\textwidth}
%	\centering
%	\includegraphics[width=\textwidth]{Figs/4lhcfpPB_100M_events/{lhcf_P_pB_9.8-10.0}.pdf}
%	\end{subfigure}
%	\begin{subfigure}{0.3\textwidth}
%	\centering
%	\includegraphics[width=\textwidth]{Figs/4lhcfpPB_100M_events/{lhcf_P_pB_10.0-10.2}.pdf}
%	\end{subfigure}
%	\begin{subfigure}{0.3\textwidth}
%	\centering
%	\includegraphics[width=\textwidth]{Figs/4lhcfpPB_100M_events/{lhcf_P_pB_10.2-10.4}.pdf}
%	\end{subfigure}
%	\begin{subfigure}{0.3\textwidth}
%	\centering
%	\includegraphics[width=\textwidth]{Figs/4lhcfpPB_100M_events/{lhcf_P_pB_10.4-10.6}.pdf}
%	\end{subfigure}
	\caption{Neutral pion transverse momentum spectra in the rapidity range $8.9<y<10.6$ in p-Pb collisions at $\sqrt{s_{NN}}=5.02\,\text{TeV}$ measured at LHCf detector. Solid and dashed lines correspond to $K$-factors $K$=1 and $K$=1.5 respectively.}
	\label{pPbLHCf}
	\end{figure}
	
	Similarly to the previous analysis presented in \cite{Deng:2014vda}, we obtain a very good description of p-p data for all rapidities, see Fig. \ref{ppLHCf}. Importantly, the $K$-factor used for the description of data is exactly the same as the one used for the description of BRAHMS data, $K=1$. This is an important result, as it indicates that the energy evolution from RHIC to LHC, equivalent to more than ten units in evolution rapidity, $\Delta Y\gtrsim 14$, is well accounted for by the theoretical tools in our approach, namely the running coupling BK evolution for the $x$-dependence of the uGDs.
For the sake of illustration in Fig. \ref{ppLHCf} we also show the partonic spectra generated prior to the hadronization process. As a comment, it should be noted that the bump observed for the lowest momentum bin is due to the contribution of projectile remnants not participating into the hard scattering.
	We also find a good agreement of the neutral pion spectra measured in p-Pb collisions, see Fig. \ref{pPbLHCf}. In this case our theoretical result is a bit below the data at the highest values of transverse momenta. Again, we have used a $K$-factor $K=1$ for its description. As shown in Fig. \ref{pPbLHCf}, a slightly larger value of the $K$-factor, $K$=1.5, results in a slightly better description of the data, although we do not have a clear motivation for such choice.

\section{Nuclear modification factor at LHCf}
Finally, in this section we present our results for the nuclear modification factor $R_{\text{p-Pb}}$, defined as follows:
	\begin{equation}\label{Rppb}
		R^{\pi^0}_{\text{p-Pb}}\equiv \frac{\sigma^{\text{pp}}_{\text{inel}}}{\langle N_{coll} \rangle \sigma^{\text{pPb}}_{\text{inel}}} \frac{Ed^{3}\sigma^{\text{pPb}}/d^3p}{Ed^{3}\sigma^{\text{pp}}/d^3p} = \frac{1}{\langle N_{coll} \rangle}\frac{dN^{\text{pPb}\rightarrow \pi^0X}/dyd^2p_t}{dN^{\text{pp}\rightarrow \pi^0X}/dyd^2p_t}.
	\end{equation}
Where $Ed^{3}\sigma^{\text{pPb}}/dp^{3}$, $Ed^{3}\sigma^{\text{pp}}/dp^{3}$ are the inclusive cross sections of neutral pion production in p-Pb and p-p collisions respectively, and $\langle N_{coll} \rangle$ is the average number of nucleon-nucleon scatterings in a p-Pb collision. We shall use the same value of $\langle N_{coll} \rangle$ as the one used in the experimental analysis \cite{Adriani:2015iwv}, obtained from a Monte Carlo Glauber simulation: $\langle N_{coll} \rangle = 6.9$. Also, it should be kept in mind that the experimental value for $\sqrt{s}\!=\!5.02$ TeV is obtained after interpolating p-p data from 2.76 and 7 TeV collision energies.

One remarkable feature of experimental data is the approximate flatness of the $R^{\pi^0}_{\text{p-Pb}}$ over all the measured rapidity range. Actually a constant value $R^{\pi^0}_{\text{p-Pb}}=1/ \langle N_{coll} \rangle \approx 0.15$ is compatible with data for all $y$. This would immediately imply that the multiplicity density in p-p collisions is approximately equal to the one in p-Pb collisions (see right hand side of \eq{Rppb}):
\begin{equation}\label{eqspec}
\frac{dN^{pp\rightarrow \pi^0X}}{dyd^2p_t}\approx \frac{dN^{pPb\rightarrow \pi^0X}}{dyd^2p_t}.
\end{equation}
%%new

\begin{figure}
\centering
\includegraphics[width=0.55\textwidth]{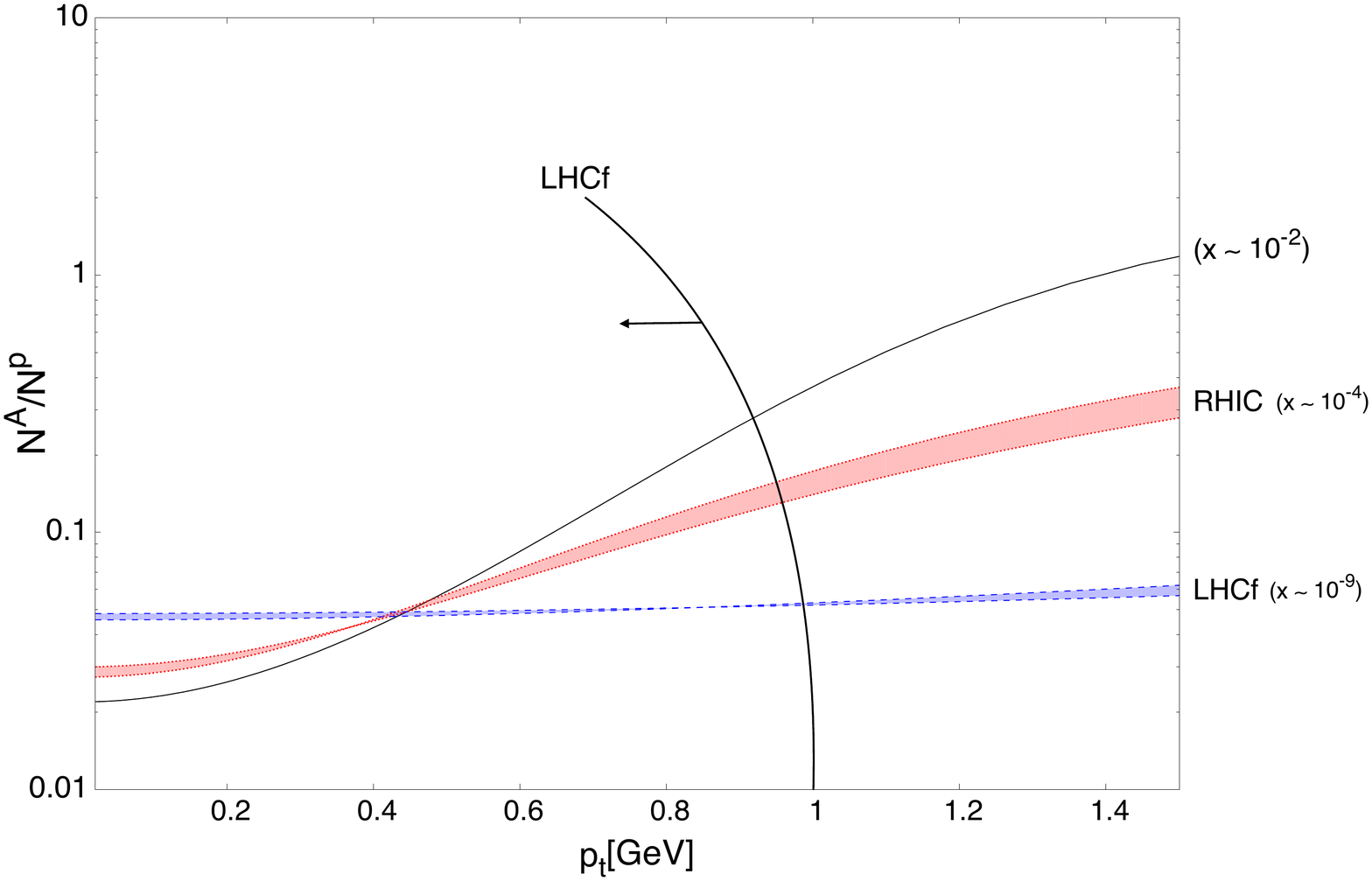}
\caption{Ratio of rcBK-evolved unintegrated gluon distributions for proton and lead targets for the different $x$-ranges observed at RHIC and LHC.}
\label{univ}
\end{figure}

Certainly, a more refined analysis of data would probably indicate a decreasing behaviour of $R^{\pi^0}_{\text{p-Pb}}$ with increasing rapidity of the detected pions. However, such rate of change is much smaller than the one observed at RHIC energies in a similar range of transverse momenta. 
This purely empirical observation is well accounted for by our calculations.
In terms of saturation physics this result can be immediately related to the asymptotic properties of the solution of the BK equation, used to describe the $x$-dependence of the uGDs of the proton and lead targets.
At partonic level, \eq{eqspec} can be written as:
\begin{equation}\label{eqspecpart}
\langle n_{\text{p-Pb}}\rangle_{b} N_{(F/A)}^{\text{Pb}} \approx \langle n_{\text{p-p}}\rangle_{b} N_{(F/A)}^{\text{p}}.
\end{equation}
Where $N_{(F/A)}^{\text{Pb}}$, $N_{(F/A)}^{\text{p}}$ are the uGDs corresponding to proton and nucleus targets, and $\langle n_{\text{p-Pb}}\rangle_{b}$, $\langle n_{\text{p-p}}\rangle_{b}$ are the average number of independent hard collisions per p-p and p-Pb events integrated in impact parameter. Due to the normalization of the spatial overlap function for proton-nucleus collisions $T_{pA}$, the integration of \eq{MPI} over $b$ yields:
\begin{equation}\label{eq}
\langle n_{\text{p-Pb}}\rangle_{b} = A^{2/3} \langle n_{\text{p-p}}\rangle_{b} .
\end{equation}
Applying this expression to \eq{eqspecpart}, and also neglecting the difference in the factorisation scales for p or Pb scattering we get:
\begin{equation}\label{eq}
\frac{N_{(F/A)}^{\text{Pb}}}{N_{(F/A)}^{\text{p}}} = \frac{1}{A^{2/3}} .
\end{equation}
This behaviour is well realised by the BK-evolved uGD's used in this work. As shown in Fig. \ref{univ}, the ratio of lead over proton uGD's takes a constant value $1/A^{2/3}\approx 0.03$ in all the $k_t$ range probed by the LHCf data studied here.
%Indeed, it is well known that deep in the saturation regime ($x\ll1$ or $\Delta Y \gg 1$) the solutions of the BK equation for the uGD's become independent of the initial conditions\cite{Albacete:2004gw}. More precisely, at high enough rapidities the solutions of the BK equation exhibit the geometric scaling property, namely they depend on a single dimensionless scale: $\mathcal{N}(x,r)\rightarrow\mathcal{N}(\tau\equiv r\,Q_s(x))$. In the scaling regime, the uGD's as given by Eq.(\ref{phihyb}) satisfy the following scaling relation:
%\begin{equation}\label{scalingU}
%N_{F,A}(x,k_t)=\frac{1}{Q_s^2(x)} N_{F,A}\left(\frac{k_t}{Q_s(x)}\right)\,.
%\end{equation}
%This feature of non-linear evolution translates into the flatness of the ratio of the uGD's corresponding to different initial conditions --i.e proton \textit{vs} nucleus-- as a function of transverse momentum, as shown in Fig. \ref{univ}.
We interpret the fact that the experimental data on nuclear modification factor reproduces this constant behaviour over the whole range of rapidity is an indication for the prevalence of saturation effects in the probed kinematic regime by the LHCf. In turn, the analogous ratio, when plotted for the kinematic regime relevant for forward RHIC data, exhibits a growing behaviour with increasing transverse momentum, in the very same fashion as the corresponding nuclear modification factor. We conclude that RHIC forward kinematics falls outside the universality regime of small-$x$ evolution. Rather, RHIC kinematics test non-linear evolution in the pre-asymptotic regime. 

%On the other hand, the shape of the curve corresponding to the kinematical region probed at RHIC exhibits a more delayed onset of saturation effects, as in this case the uGD of the target is characterized by a smaller value for the saturation scale.
%%

	\begin{figure}[htbp]
	\centering
	
		\subfigure{\label{fig:6a}\includegraphics[width=0.3\textwidth]{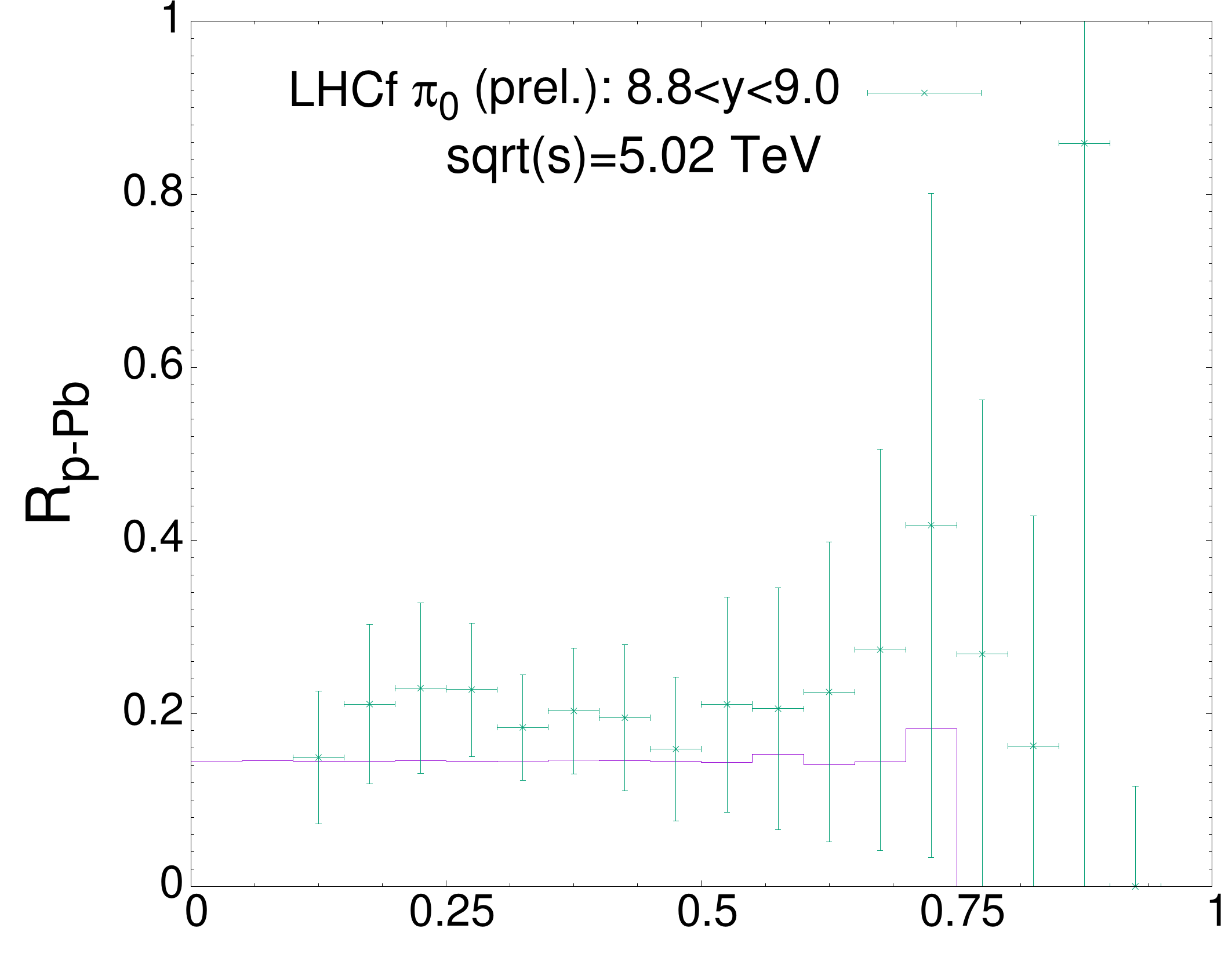}}
		\subfigure{\label{fig:6b}\includegraphics[width=0.3\textwidth]{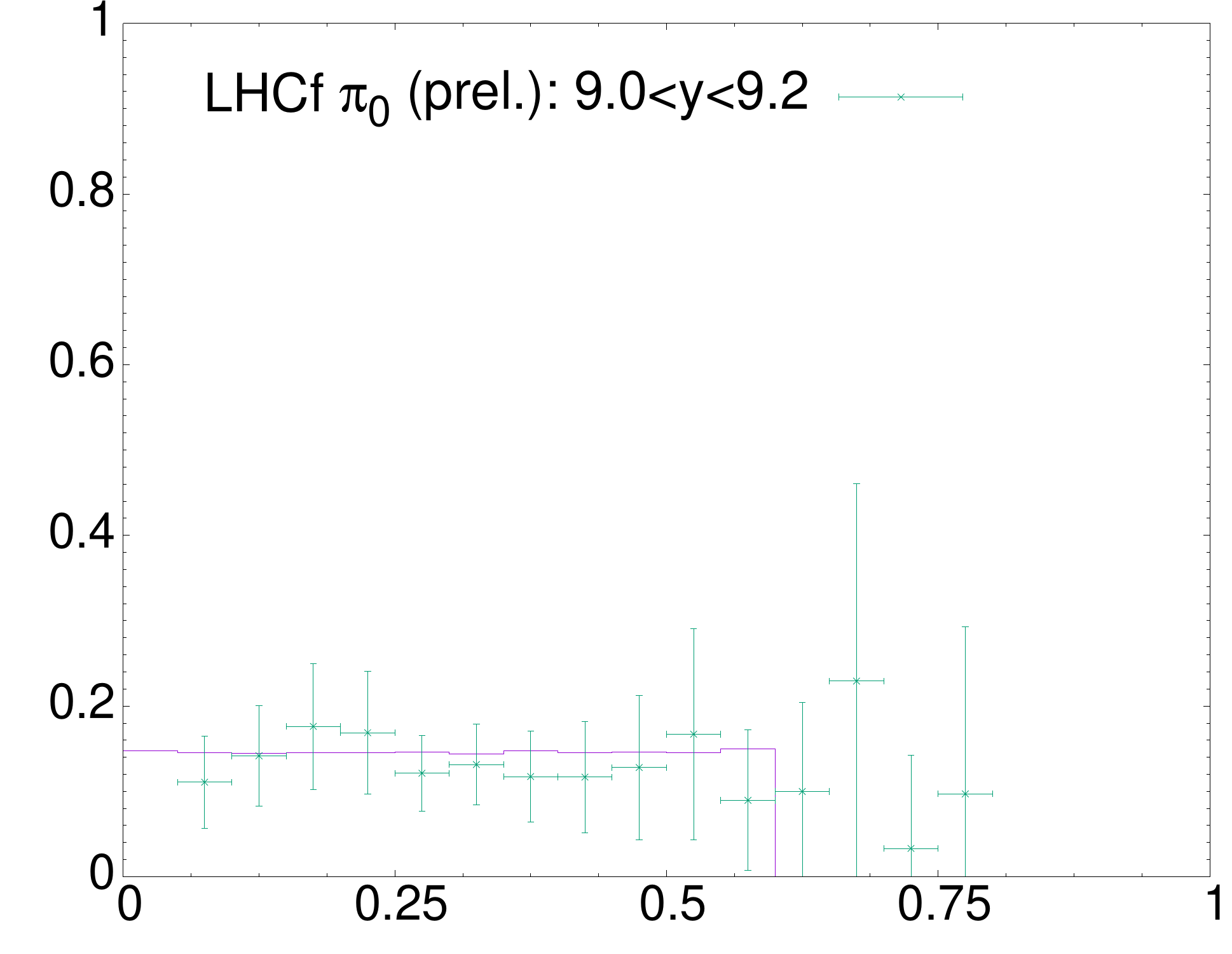}}
		\subfigure{\label{fig:6c}\includegraphics[width=0.3\textwidth]{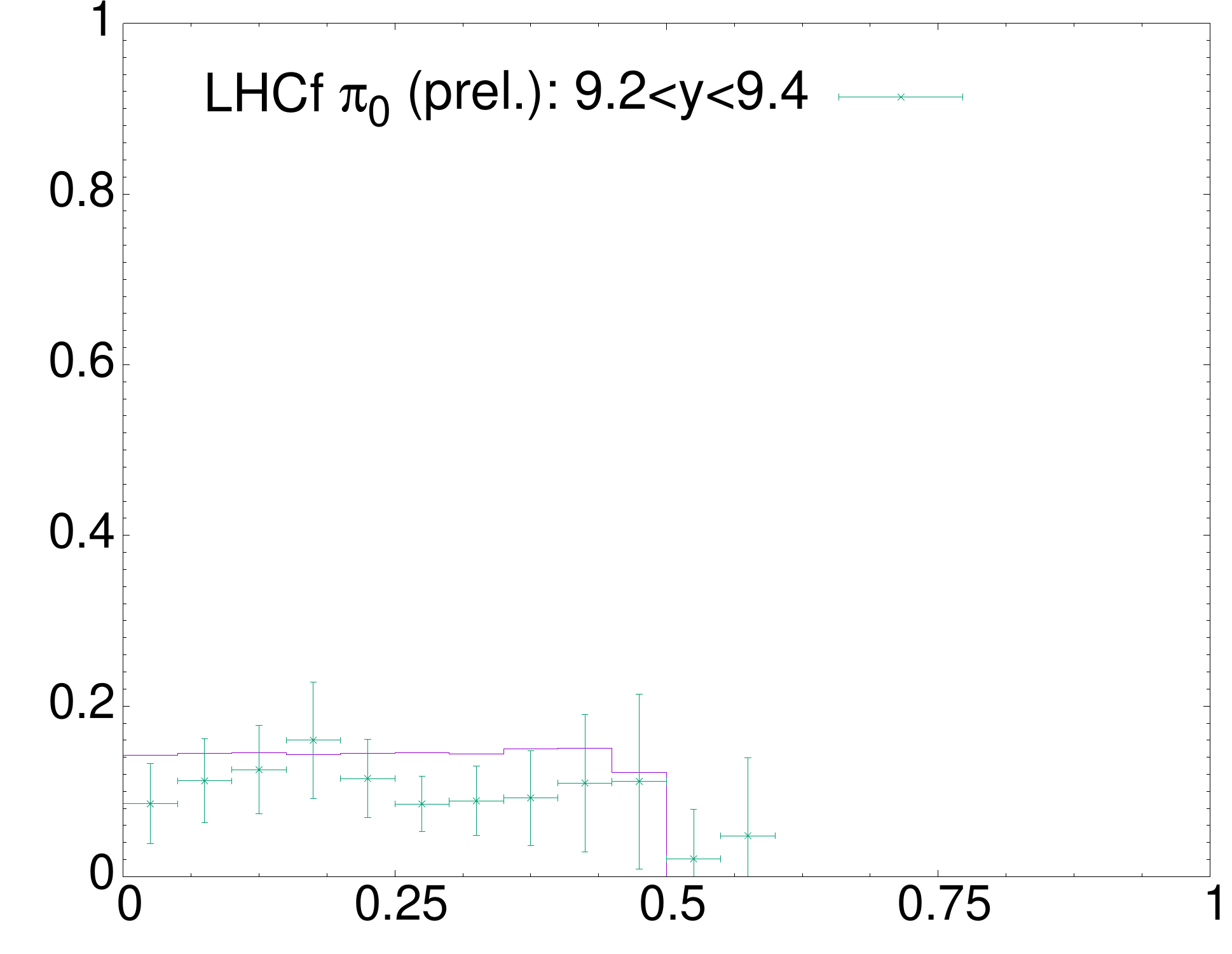}}
		\subfigure{\label{fig:6d}\includegraphics[width=0.3\textwidth]{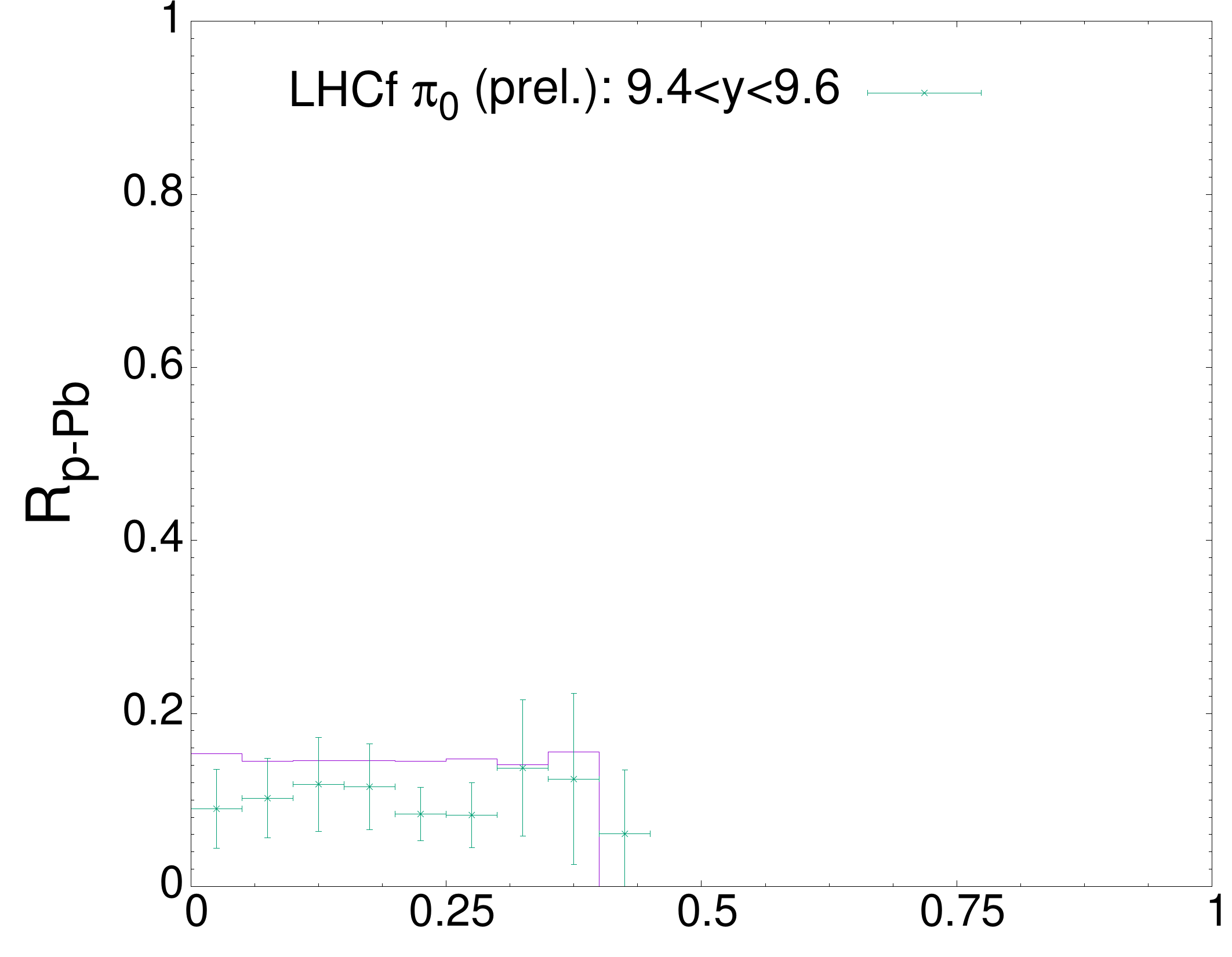}}
		\subfigure{\label{fig:6e}\includegraphics[width=0.3\textwidth]{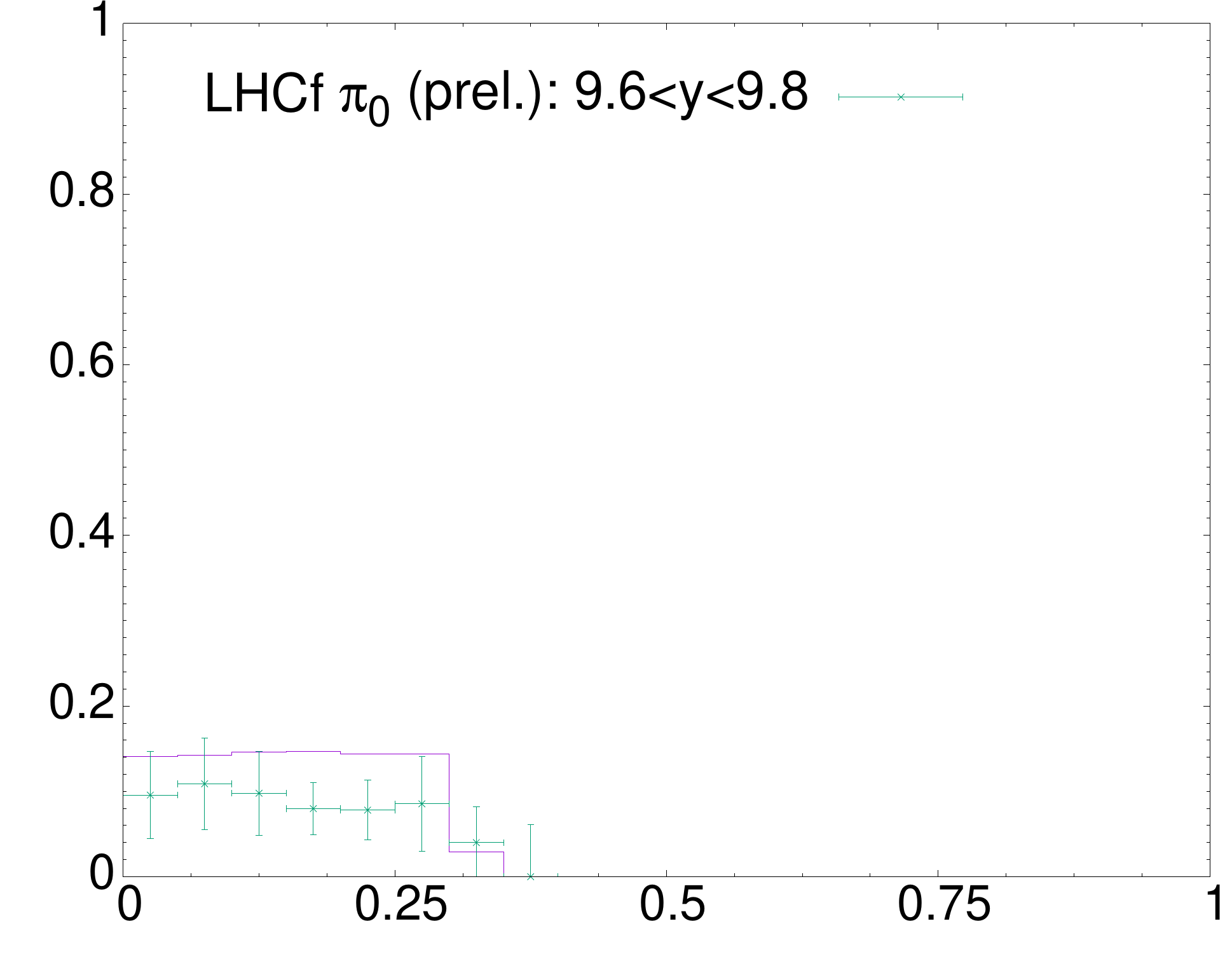}}
		\subfigure{\label{fig:6f}\includegraphics[width=0.3\textwidth]{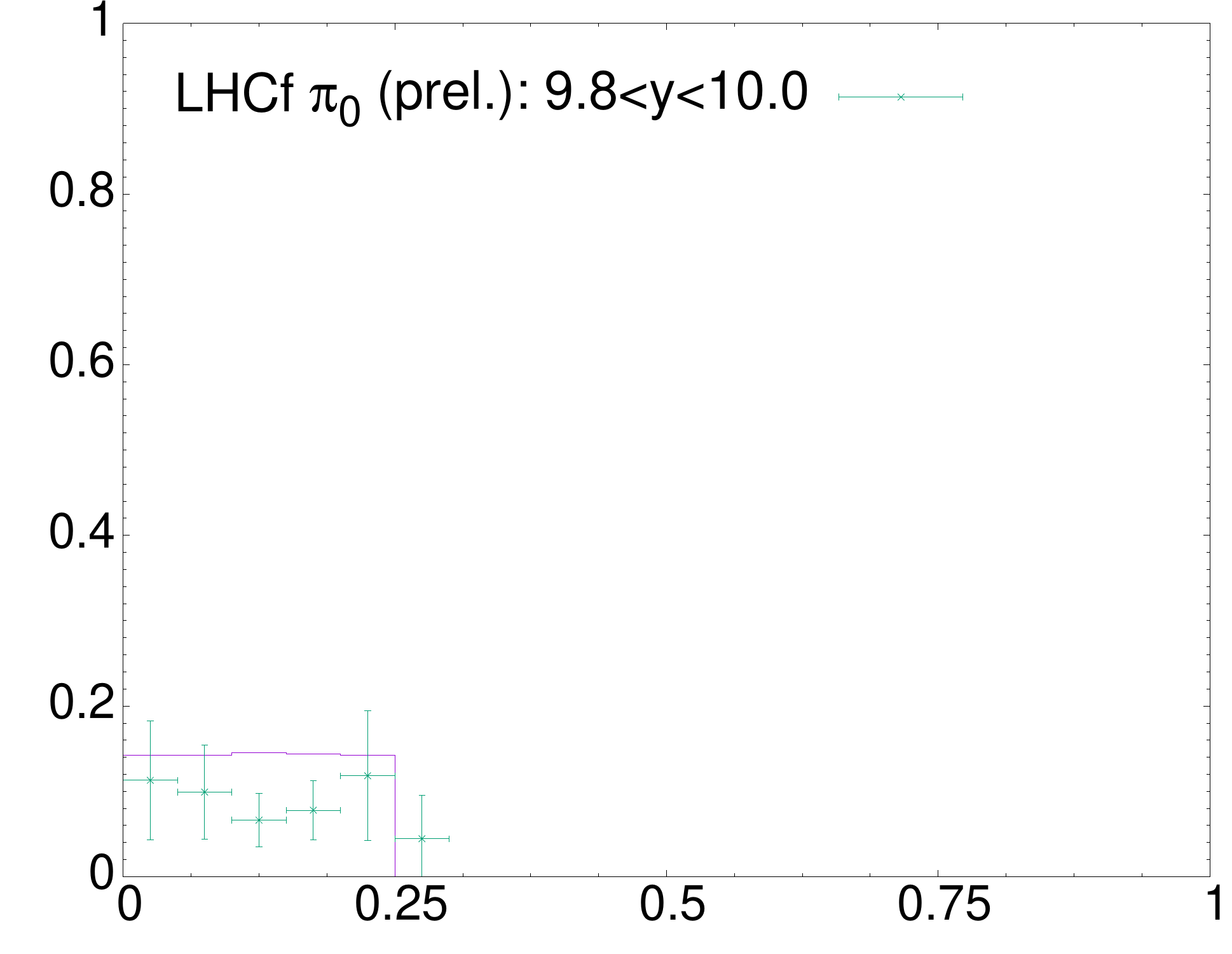}}
		\subfigure{\label{fig:6g}\includegraphics[width=0.3\textwidth]{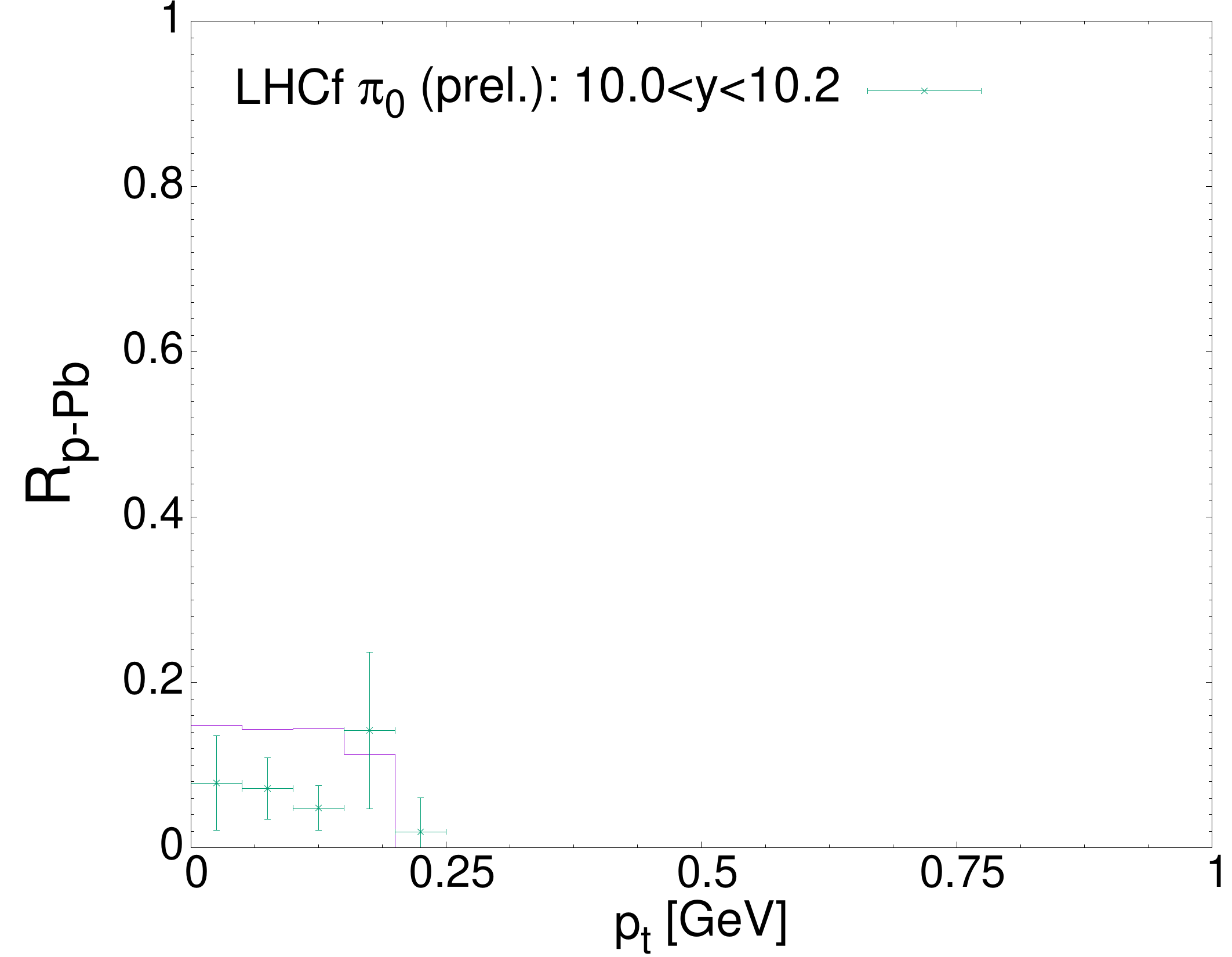}}
		\subfigure{\label{fig:6h}\includegraphics[width=0.3\textwidth]{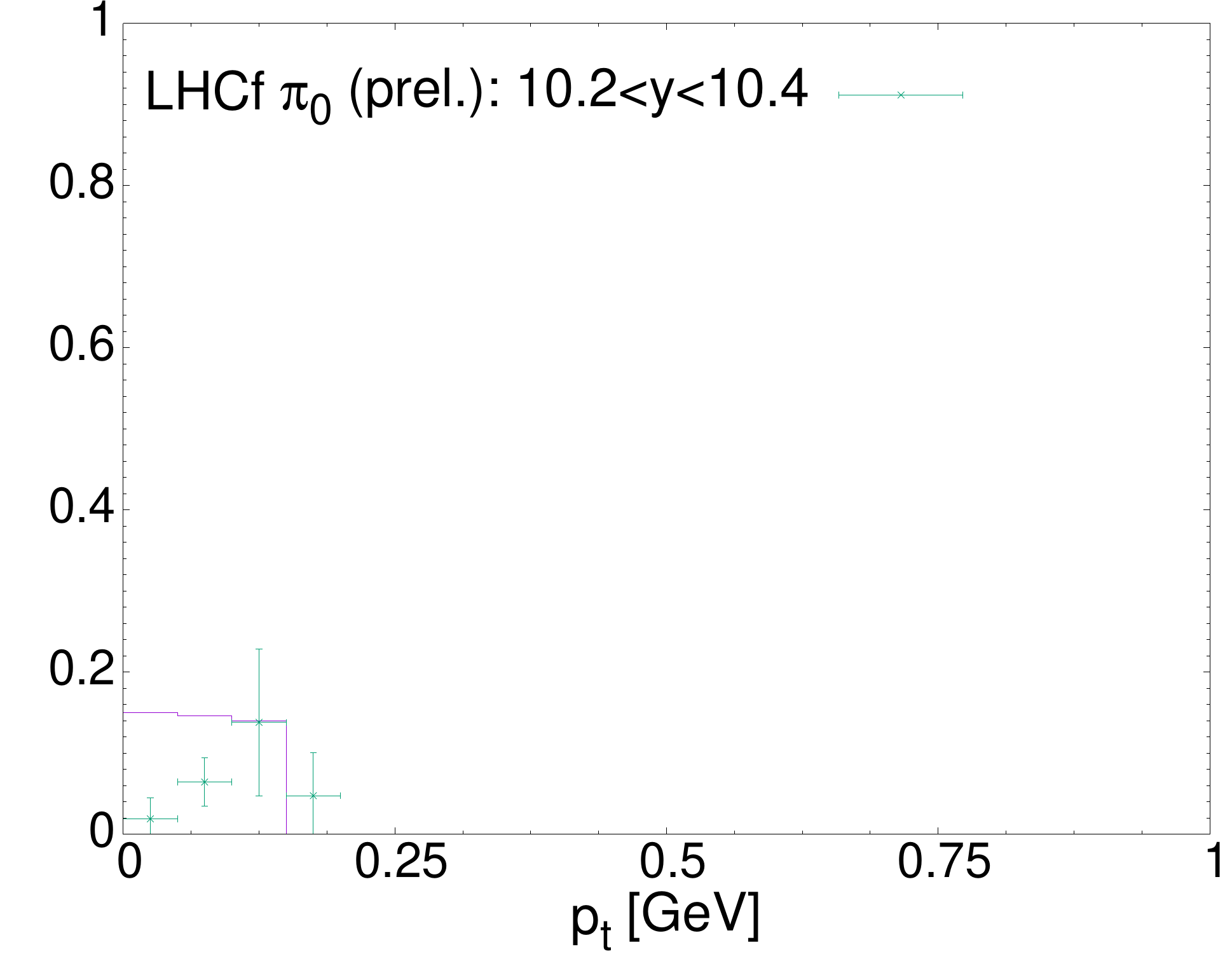}}
		\subfigure{\label{fig:6i}\includegraphics[width=0.3\textwidth]{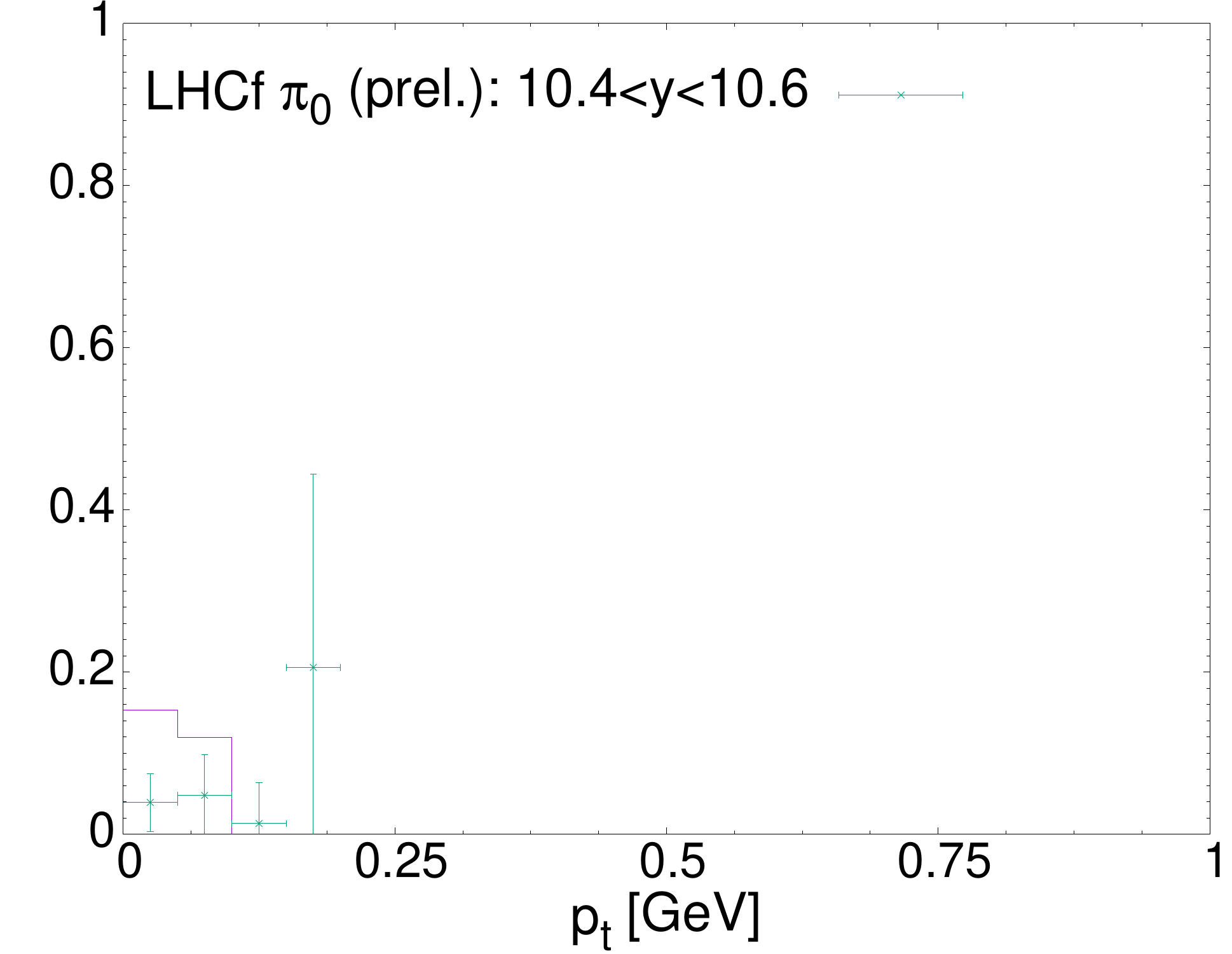}}	
	
%	\begin{subfigure}{0.3\textwidth}
%	\centering
%	\includegraphics[width=\textwidth]{Figs/5lhcf_modification_factor_100M_events/{lhcf_mf_8.8-9.0}.pdf}
%	\end{subfigure}
%	\begin{subfigure}{0.3\textwidth}
%	\centering
%	\includegraphics[width=\textwidth]{Figs/5lhcf_modification_factor_100M_events/{lhcf_mf_9.0-9.2}.pdf}
%	\end{subfigure}
%	\begin{subfigure}{0.3\textwidth}
%	\centering
%	\includegraphics[width=\textwidth]{Figs/5lhcf_modification_factor_100M_events/{lhcf_mf_9.2-9.4}.pdf}
%	\end{subfigure}
%	\begin{subfigure}{0.3\textwidth}
%	\centering
%	\includegraphics[width=\textwidth]{Figs/5lhcf_modification_factor_100M_events/{lhcf_mf_9.4-9.6}.pdf}
%	\end{subfigure}
%	\begin{subfigure}{0.3\textwidth}
%	\centering
%	\includegraphics[width=\textwidth]{Figs/5lhcf_modification_factor_100M_events/{lhcf_mf_9.6-9.8}.pdf}
%	\end{subfigure}
%	\begin{subfigure}{0.3\textwidth}
%	\centering
%	\includegraphics[width=\textwidth]{Figs/5lhcf_modification_factor_100M_events/{lhcf_mf_9.8-10.0}.pdf}
%	\end{subfigure}
%	\begin{subfigure}{0.3\textwidth}
%	\centering
%	\includegraphics[width=\textwidth]{Figs/5lhcf_modification_factor_100M_events/{lhcf_mf_10.0-10.2}.pdf}
%	\end{subfigure}
%	\begin{subfigure}{0.3\textwidth}
%	\centering
%	\includegraphics[width=\textwidth]{Figs/5lhcf_modification_factor_100M_events/{lhcf_mf_10.2-10.4}.pdf}
%	\end{subfigure}
%	\begin{subfigure}{0.3\textwidth}
%	\centering
%	\includegraphics[width=\textwidth]{Figs/5lhcf_modification_factor_100M_events/{lhcf_mf_10.4-10.6}.pdf}
%	\end{subfigure}
	\caption{Nuclear modification factor for neutral pion production at $\sqrt{s_{NN}}=5.02\, \text{TeV}$. Data points taken from \cite{Adriani:2015iwv}. Since there is no neutral pion transverse momentum spectra measurement available for p-p collisions at $\sqrt{s}=5.02\, \text{TeV}$, it is derived by interpolation of datasets obtained from p-p collisions at $\sqrt{s}=7\, \text{TeV}$ and $2.76\, \text{TeV}$, which are included in that paper.}
	\label{ModFac}
	\end{figure}
%To compute the theoretical curve in figure \ref{ModFac} we generated 100 million p-p and p-Pb events and calculated (\ref{Rppb}) for every transverse momentum bin. The large number of events was necessary to suppress statistical fluctuations, which have a significant distorting effect on $R_{\text{p-Pb}}$. The result obtained for all rapidity ranges is a flat plateau at $R_{\text{p-Pb}}=0.15$, a value that roughly matches the inverse of $\langle N_{coll} \rangle =6.9$. Thus, figure \ref{ModFac} shows that experimental data is compatible with $Ed^{3}\sigma^{\text{pPb}}/dp^{3}=Ed^{3}\sigma^{\text{pp}}/dp^{3}$. This may indicate that the saturation momentum is well above the measured $p_{T}$ range for both p-p and p-Pb collisions, resulting on a universality in particle production.

\section{Discussion}

In this work we have shown that it is possible to obtain a good description of particle production (single neutral pion spectrum) in the very forward region of the LHC and down to the lowest values of transverse momentum experimentally accessed by the LHCf collaboration ($p_t\lesssim 0.1$ GeV). We do so via the combination of a perturbative description of the elementary partonic scattering process with a non-perturbative \textit{stringy} characterization of the fragmentation and decay of the \textit{hard} partons. Our results provide yet another indication for the presence --and need-- of saturation effects to correctly describe presently available experimental data dominated by the contribution of very small-$x$ gluons. 
Indeed, the good and simultaneous description of p-p and p-Pb data and, in particular, of the nuclear modification $R_{pPb}$ factor affords a neat theoretical interpretation in terms of saturation physics. The flatness and approximate constant behaviour of the observed $R_{pPb}$ over the wide range of rapidities covered by data can be related to the asymptotic properties of the solutions of the BK equation and, in particular, to the existence of universal solutions at sufficiently small-$x$.
%... , as the only difference between the uGD's for a proton and a nucleus is a rescaling of the initial saturation scale in \eq{Init} (by the \textit{oomph} factor $Q^2_{s0,nucleus}=A^{1/3}Q_{s0,proton}^2$), in the very small-$x$ limit both uGD's show an asymptotically similar behaviour. (...)

Aside from the description of the data discussed in this work, the fact that the main features of ultra forward production data --even for very small transverse momentum of the produced particles-- can be understood in terms of perturbative tools may open interesting new avenues of research in the field of ultra high-energy cosmic rays (UHECR). There, the main features of the air showers developed after the primary collisions in the upper atmosphere are determined to a large extent by the hadronic collisions properties, in particular, by the total cross-section, forward multiplicity, charm production, inelasticity, etc. \cite{Ulrich:2009hm}. Thus, the availability of theoretically controlled tools to extrapolate from the well constrained collision energy domain probed at the LHC to that of UHECR is necessary to reduce the inherent uncertainty associated to the extrapolation itself and, thereby, also the uncertainty associated to the present analysis of the primary mass composition of UHECR. We propose that the use of non-linear renormalization group equations of QCD (like the BK equation employed in this work) can offer insight in this direction and we plan to extend our studies in this direction in future works.

%In this paper we use a Monte-Carlo implementation of the DHJ formula combined with Lund fragmentation to perform an analysis of data on ultra-forward inclusive hadron production in high-energy proton-proton and proton-nucleus collisions in RHIC and LHC. We assume the forward rapidity regions probed by these collisions is sensible to the onset of gluon saturation effects on the target hadron or nucleus. This nonlinear dynamics is controlled by a perturbatively large momentum scale $Q_{s}$. $\alpha(Q_{s}^{2})\ll 1$ makes it possible to perform a perturbative approach. The combination with the Lund fragmentation model for hadronization provides a unified description of hadron spectra from low to high momentum region that consistently reproduces experimental data at RHIC and LHC energies. Uncertainties of the model are accounted for with a $K-$factor unaffected by nuclear effects. Our results on nuclear modification factors suggest that ultra-forward hadron production in p-Pb collisions at $\sqrt{s}=5.02\,\text{TeV}$ is profoundly dominated by gluon saturation. We are planning to extend this approach to nucleus-nucleus collisions by use of nuclear PDF's (nPDF's) in the near future. In the not-so-distant future, we expect to apply our results to the calculation of charged particle multiplicities in the ultra-forward rapidity regime. This would serve as a good startpoint for a study of the mass composition of ultra-high energy cosmic rays, as forward particle production is of key importance in the development of air showers.

\bibliographystyle{elsarticle-num}
%\clearpage
%\bibliography{ref2}{}

%\end

\end{document}